\documentclass{aastex63}

\usepackage{newtxmath}
\usepackage{graphicx}
\usepackage{caption}
\usepackage[skip=0.5ex]{subcaption}

\newcommand{\prior}{\textit{a priori}}

\newcommand{\amoeba}{\textsc{Amoeba}}
\newcommand{\emcee}{\textsc{Emcee}}

\received{June 1, 2019}
\revised{January 10, 2019}
\accepted{\today}
\submitjournal{ApJ}

\shorttitle{\amoeba}
\shortauthors{Petzler et al.}
\graphicspath{{./}{Figures/}}

\begin{document}

\title{\amoeba: Automated Molecular Excitation Bayesian Line-Fitting Algorithm}

\correspondingauthor{Anita Petzler}
\email{anita.petzler@students.mq.edu.au}

\author[0000-0001-6179-0606]{Anita Petzler}
\affiliation{Department of Physics and Astronomy, Macquarie University, NSW 2109, Australia}
\affiliation{Research Centre in Astronomy, Astrophysics and Astrophotonics, Macquarie University, NSW 2109, Australia}

\author{J. R. Dawson}
\affiliation{Department of Physics and Astronomy, Macquarie University, NSW 2109, Australia}
\affiliation{Research Centre in Astronomy,  Astrophysics and Astrophotonics, Macquarie University, NSW 2109, Australia}
\affiliation{CSIRO Astronomy and Space Science, Australia Telescope National Facility, PO Box 76, Epping, NSW 1710, Australia}

\author{Mark Wardle}
\affiliation{Department of Physics and Astronomy, Macquarie University, NSW 2109, Australia}
\affiliation{Research Centre in Astronomy, Astrophysics and Astrophotonics, Macquarie University, NSW 2109, Australia}

\begin{abstract}

The hyperfine transitions of the ground-rotational state of the hydroxyl radical (OH) have emerged as a versatile tracer of the diffuse molecular interstellar medium. We present a novel automated Gaussian decomposition algorithm designed specifically for the analysis of the paired on-source and off-source optical depth and emission spectra of these transitions. In contrast to existing automated Gaussian decomposition algorithms, \amoeba~(Automated MOlecular Excitation Bayesian line-fitting Algorithm) employs a Bayesian approach to model selection, fitting all 4 optical depth and 4 emission spectra simultaneously. \amoeba~assumes that a given spectral feature can be described by a single centroid velocity and full width at half-maximum, with peak values in the individual optical depth and emission spectra then described uniquely by the column density in each of the four levels of the ground-rotational state, thus naturally including the real physical constraints on these parameters. Additionally, the Bayesian approach includes informed priors on individual parameters which the user can modify to suit different data sets. Here we describe \amoeba~and evaluate its validity and reliability in identifying and fitting synthetic spectra with known parameters.

\end{abstract}


\section{Introduction}\label{sec:intro}
    The gas and dust between the stars -- the Galactic interstellar medium (ISM) -- plays an important role in the dynamic evolution of our Galaxy and others. The molecular gas of the ISM (the vast majority of which is H$_2$) is key to several important stages of that evolution. Unfortunately, molecular hydrogen does not have readily accessible energy levels at the low temperatures and densities seen in the majority of the molecular ISM, rendering it effectively invisible. Consequently, other molecules with accessible states in this low-energy environment that are expected to coexist with ${\rm H}_2$ are instead observed as tracers of the molecular gas. The hydroxyl radical (OH) is one such tracer. 

\begin{figure}
    \centering
    \includegraphics[width=\linewidth]{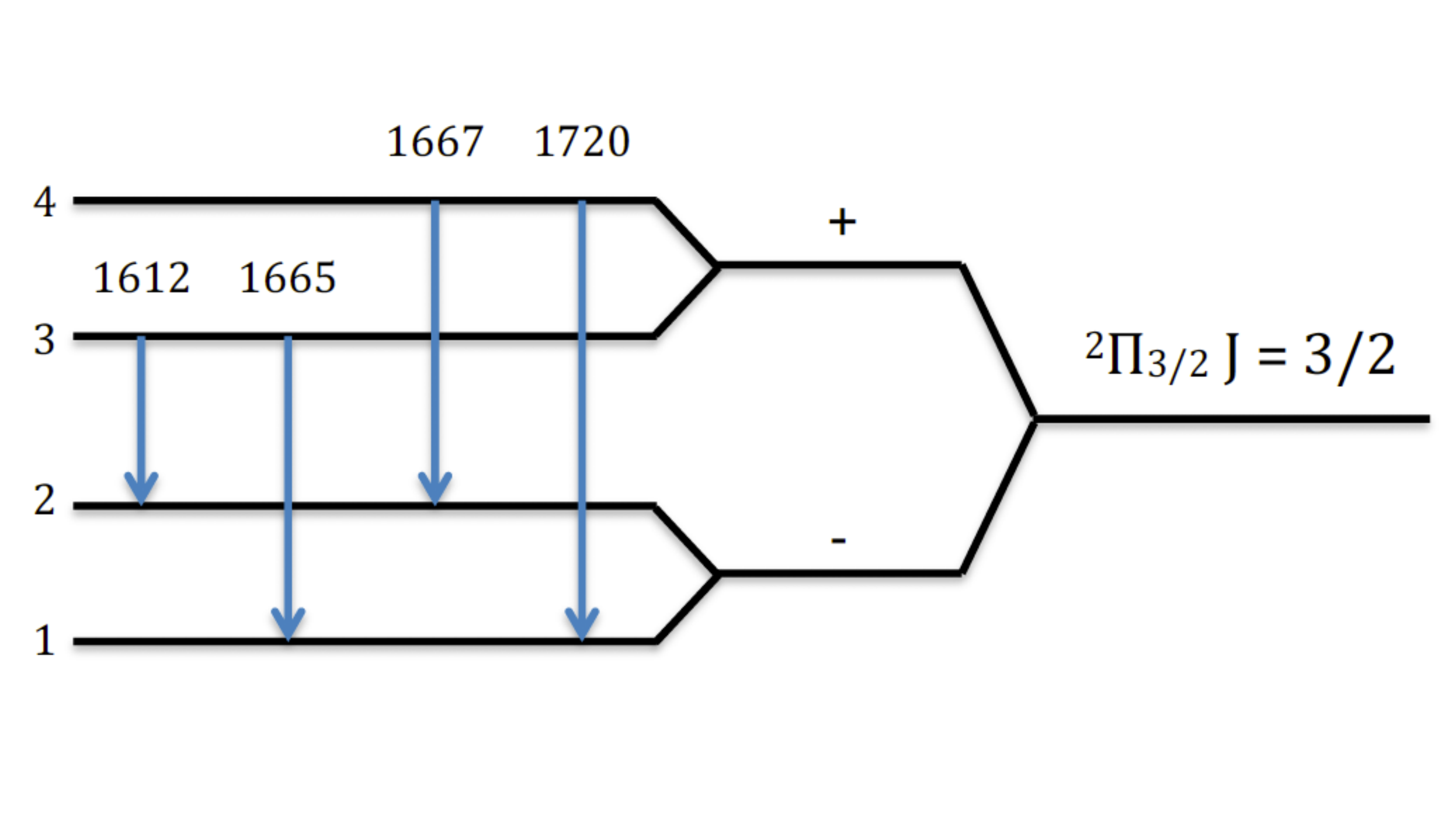}
    \caption{The ground rotational state of the hydroxyl radical and hyperfine transitions at 1612, 1665, 1667 and 1720 MHz. The levels are numbered 1-4 from lowest to highest energy.}
    \label{fig:OH_ground}
\end{figure}

The ground-rotational state of OH is split into 4 levels by $\Lambda$-doubling and hyperfine splitting, with 4 allowed transitions between those levels at 1612.231, 1665.402, 1667.359 and 1720.530 MHz. These levels and transitions are illustrated in Figure \ref{fig:OH_ground}. OH was the first molecule discovered in the ISM \citep{Weinreb1963}, and observations of the hyperfine transitions within its ground-rotational state have been used extensively to trace the distribution, properties and dynamics of ISM gas \citep[e.g.][]{Robinson1967,Goss1968,Turner1979,Liszt1996,Li2018,Rugel2018,Busch2019,Heiles1969,Ebisawa2015,Ebisawa2019,Petzler2020}. Crucially, OH has been demonstrated both theoretically \citep{Wolfire2010} and through observations \citep{Barriault2010} to correlate with the molecular hydrogen of the ISM across a wide range of number densities and column densities \citep[see also][and references therein]{Nguyen2018}. This link has led to an increase in the use of OH as a supplementary tracer of molecular hydrogen \citep[e.g.][]{Dawson2014,Nguyen2018,Li2018,Rugel2018,Engelke2019}, particularly in diffuse regions \citep{Wannier1993,Liszt1996,Barriault2010,Allen2012,Allen2015}. This is in contrast to observations of the low rotational transitions of CO, which have been shown to under-estimate the mass of molecular gas in diffuse regions \citep[e.g.][]{Reach1994,Grenier2005,Planck2011,Remy2018}.

A key advantage of observing OH is the sensitivity of its ground-rotational state level populations to local gas conditions such as kinetic temperature, number density and radiation field \citep{Guibert1978,Elitzur1976a,Elitzur1976b,Petzler2020}. Therefore by measuring the distribution of OH molecules across the four levels of the ground-rotational state, non-LTE molecular excitation modelling can then be used to constrain these local conditions \citep{Ebisawa2015,Ebisawa2019,Petzler2020} which in turn can inform a wide range of theories of ISM phenomena. Though this distribution can in principle be determined directly from observable quantities for a single ISM cloud along a given line of sight, most observations will capture several Doppler-shifted and potentially blended spectral features representing several distinct ISM clouds. This blending necessitates an initial step where these blended features are decomposed. However, this presents several potential complications of relevance to this work. Chief among these is the question of the number and velocity position of individual components. As outlined in Section \ref{Gauss_decomp} and discussed in detail in Section \ref{Sec:Amoeba}, our approach calculates Bayes factors of competing Gaussian decomposition models to find the `best' decomposition (given the \textit{a priori} and likelihood distributions). Another key question regards the treatment of the four transitions during this process of Gaussian decomposition: namely, whether the spectra should be decomposed independently or if they should be linked in some way. As we discuss in Section \ref{Sec:Amoeba}, we have chosen the novel approach to decompose all four spectra of the ground-rotational state transitions simultaneously. We take advantage of the intrinsic relationships between the parameters of the Gaussian-shaped features in each of the four transitions' spectra (i.e. centroid velocity, full width at half-maximum and peak values in each transition) and the underlying excitation properties of the OH molecules (i.e. the column densities in each level of the ground-rotational state). This approach allows us to take greater advantage of the available data: by using each transition as additional information about the same gas, we are able to focus only on models of the OH excitation that are completely self-consistent to the limit of our underlying assumptions (discussed in depth in Section \ref{Sec:Amoeba}).

In this paper we introduce \amoeba: an open-source Automated Molecular Excitation Bayesian line-fitting Algorithm for the simultaneous decomposition of all four OH ground-rotational spectra written in \textsc{Python}. \amoeba~is intended to be used for large OH data sets, such as the H\textsc{i}/OH/Recombination line survey of the Milky Way \citep[THOR][]{Beuther2016}, or the upcoming Galactic Australian Square Kilometre Array Pathfinder Survey \citep[GASKAP][]{Dickey2013}. These large-scale surveys would take full advantage of \amoeba's automated nature, in that it takes spectra as an input and outputs fully decomposed spectra fit with scientifically useful parameters that can then be used with molecular excitation modelling to determine or constrain the local conditions of the gas.

In Section \ref{Sec:Background} we discuss the OH ground-rotational state and the observable quantities related to the distribution of molecules across the levels of that state. Section \ref{Sec:Background} also discusses the place of this work in the context of other Gaussian decomposition algorithms. In Section \ref{Sec:Amoeba} we introduce and describe \amoeba, and in Section \ref{Sec:Discussion} we describe its performance in a series of tests on synthetic and real data. In Section \ref{Sec:Conclusions} we conclude.

\section{Background}\label{Sec:Background}

\subsection{Observable quantities}

If gas containing OH molecules lies between the observer and a compact background source of continuum with brightness temperature $T_{\rm c}$~and the spatially extended background continuum (due to Galactic synchrotron and the cosmic microwave background) $T_{\rm bg}$, the observed continuum-subtracted line brightness temperature is described by:
    \begin{equation}
    T_{\rm b} = (T_{\rm ex}-T_{\rm c}-T_{\rm bg})(1-e^{-\tau_{\nu}}),
    \label{Tb}
    \end{equation}
\noindent where the excitation temperature $T_{\rm ex}$~is a reparameterisation of the population of the upper and lower levels of the given transition within the cloud, defined by the Boltzmann factor:
    \begin{equation}
    \frac{N_u}{N_l} = \frac{g_u}{g_l} \exp{\bigg[\frac{-h\nu_0}{k_{\rm B} T_{\rm ex}}\bigg]},
    \label{Tex}
    \end{equation}
\noindent where $N_u$~and $N_l$~are the column densities in the upper and lower levels, respectively, $g_u$~and $g_l$~are the
degeneracies of the upper and lower levels, respectively, and $\nu_0$~is the rest frequency of the transition. The excitation temperatures of the four ground-rotational transitions are not independent, and are related via the excitation temperature sum rule:
    \begin{equation}
    \frac{\nu_{1612}}{T_{\rm ex}(1612)}+\frac{\nu_{1720}}{T_{\rm ex}(1720)}=\frac{\nu_{1665}}{T_{\rm ex}(1665)}+\frac{\nu_{1667}}{T_{\rm ex}(1667)},
    \label{Texsum}
    \end{equation}
\noindent which is derived from the definition of excitation temperature (Eq. \ref{Tex}) and the energy difference between the four levels and will therefore hold in all environments. The optical depth $\tau_{\nu}$~of an isothermal homogeneous cloud is given by:
    \begin{equation}
    \tau_{\nu}=\frac{c^2}{8\pi \nu^2_0} \frac{g_u}{g_l}\,N_l\,A_{ul} \bigg(1-\exp{\bigg[\frac{-h\nu_0}{k_{\mathrm{B}} T_\mathrm{ex}}\bigg]}\bigg)\,\phi(\nu),
    \label{tau}
    \end{equation}
\noindent where $\phi(\nu)$~is the line profile. In environments where $|T_{\rm ex}|\gg h\nu_0/k_{\rm B}=0.08$~the bracketed term in Eq. \ref{tau} is well-represented by $h \nu_0/k_B$, and the peak optical depths in the four ground-rotational transitions are then not fully independent but are related via the optical depth sum rule \citep{Robinson1967}:
    \begin{equation}
    \tau_{\rm peak}(1612) + \tau_{\rm peak}(1720) = \frac{\tau_{\rm peak}(1665)}{5} + \frac{\tau_{\rm peak}(1667)}{9}.
    \label{sum}
    \end{equation}

The shape of the line profile $\phi(\nu)$~will be determined by the dominant broadening mechanisms within the cloud, which in the molecular ISM is assumed to be turbulence. We further assume that this turbulence is on a small scale which would result in a Gaussian-shaped profile via the central limit theorem. Thus far we have assumed that a given gas cloud can be accurately described by a single $T_{\rm ex}$~value in each OH transition, despite the fact that ISM clouds are known to have clumpy and turbulent structure on all observable scales \citep[e.g.][]{Hennebelle2008,Benincasa2013}. This assumption implies that these $T_{\rm ex}$~and $N_{\rm OH}$~values -- which would in truth only describe a completely homogeneous cloud -- actually represent a quasi- `averaging' of this ensemble of smaller structures. Therefore when we ascribe a single set of $T_{\rm ex}$~and $N_{\rm OH}$~values to a Gaussian feature -- and by extension, to a ISM cloud along the line of sight -- we acknowledge that this is a simplification.

$T_{\rm ex}$~and $\tau_{\nu}$~are uniquely described by the column densities in each level of the ground-rotational state $N_1$--$N_4$, which in turn are determined by the properties and local conditions of the cloud. There are, however, two key complications in characterising $N_1$--$N_4$. First, Eq. \ref{Tb} implies that $T_{\rm ex}$~and $\tau_{\nu}$~cannot both be uniquely characterised from a single observation. Second, it is very common for multiple clouds to lie along a single line of sight, so the observed $T_{\rm b}$~at a given spectral channel may represent flux from more than one cloud, which must be disentangled before $T_{\rm ex}$~and $\tau_{\nu}$~can be determined. The second complication is readily resolved by decomposing the observed spectra into Gaussian components, then relating the parameters of those Gaussians to the target quantities $T_{\rm ex}$~and $\tau_{\nu}$, and therefore $N_1$--$N_4$.

The first complication can be solved through choice of observing strategy, two examples of which are relevant to this work. First, a single-dish telescope with a sufficiently narrow beam (e.g. Arecibo) can observe both directly `on' a background continuum source $T_{\rm b}^{\rm on}=(T_{\rm ex}-T_{\rm c}-T_{\rm bg})(1-e^{-\tau_{\nu}})$~and several `off' continuum points surrounding the `on' position $T_{\rm b}^{\rm off}=(T_{\rm ex}-T_{\rm bg})(1-e^{-\tau_{\nu}})$ \citep[e.g.][]{Heiles2003,Li2018,Nguyen2018,Nguyen2019}. This approach assumes that $T_{\rm ex}$, $T_{\rm bg}$~and $\tau_{\nu}$~are the same in the `on' and `off' positions, and also that none of the compact background continuum is present in the `off' positions. Such observations can then be converted to optical depth spectra:
\begin{equation}
    \frac{T_{\rm b}^{\rm on}-T_{\rm b}^{\rm off}}{T_{\rm c}}=e^{-\tau_{\nu}}-1.
\end{equation}

In this case the `off-source' observations then represent the brightness temperature that would be expected if the compact background continuum source could be switched off. However, in the case where the compact background continuum source is resolved and overlaps with the `off' source positions (e.g. observations towards radio-bright H\textsc{ii} regions within the Galaxy), the `on' source brightness temperature is described by $T_{\rm b}^{\rm on} =(T_{\rm ex}-T_{\rm c}^{\rm on} -T_{\rm bg})(1-e^{-\tau_{\nu}})$~and the `off' source brightness temperature is $T_{\rm b}^{\rm off}=(T_{\rm ex}-T_{\rm c}^{\rm off}-T_{\rm bg})(1-e^{-\tau_{\nu}})$. The optical depth spectra are then obtained from: 
\begin{equation}
    \frac{T_{\rm b}^{\rm on}-T_{\rm b}^{\rm off}}{T_{\rm c}^{\rm on}-T_{\rm c}^{\rm off}}=e^{-\tau_{\nu}}-1.
\end{equation}
In practice the value of $T_{\rm c}^{\rm off}$~would likely differ between `off' source positions, requiring care when generating optical depth spectra in this case.

In both cases, following the convention of \citet{Heiles2003} we can define an `expected brightness temperature' $T_{\rm exp}$~as the spectrum expected if the compact background continuum source could be switched off:
\begin{equation}\label{Texp}
    T_{\rm exp}=(T_{\rm ex}-T_{\rm bg})(1-e^{-\tau_{\nu}}),
\end{equation}
\noindent from which $T_{\rm ex}$~and therefore $N_1$--$N_4$~can be determined uniquely. \amoeba~has been designed to take as its primary input a set of these optical depth ($\tau_{\nu}$) and expected brightness temperature vs. velocity spectra. Such data sets are referred to as `on-off spectra' in this paper.

Additionally, if a bright background continuum source is observed with an interferometer, the $T_{\rm ex}$~and $T_{\rm bg}$~terms in Eq. \ref{Tb} become insignificant. This assumes that either the OH cloud (and hence the emission from it) is larger on the sky than the interferometer fringes and is therefore not detected due to spatial filtering, or that $T_{\rm c}\ll T_{\rm ex}$. (Where this latter condition may also be fulfilled for some single dish observations with a sufficiently small beam and sufficiently bright background source.) This then allows the measured $T_{\rm b}$~spectra to be directly converted to optical depth spectra. Though this method does not allow the excitation temperatures of the four transitions and therefore the  populations of the levels of the ground-rotational state to be determined directly, it does allow for useful constraints to be placed on these populations \citep{Petzler2020}. \amoeba~has therefore also been designed to also fit a set of optical depth spectra only.

\subsection{Gaussian Decomposition}\label{Gauss_decomp}

The prime difficulty in Gaussian decomposition is the question of the number and location of Gaussian components present in a spectrum. Often the human eye and informed personal judgement are relied upon to determine the number and position of Gaussian components, then a least-squares minimisation technique is employed to optimise the parameters of those Gaussians \citep[e.g.][etc.]{Heiles2003,Rugel2018,Li2018}. 
Several approaches exist to automate the process of identifying the number and position of Gaussian features, mostly in the context of the notoriously blended spectra of H\textsc{i}. One such method is that described by \citet{Haud2000}, where first a single Gaussian is placed and optimised, and then another, etc. until a threshold residual is reached. That number of components is then reduced by removing features deemed to be unreasonable or merging close features.
 
Other methods take a more analytical approach \citep[as opposed to the simple sequential approach of][]{Haud2000} to determine the number and positions of features, such as \textsc{GaussPy}~\citep{Lindner2015}~which identifies the locations of potential Gaussian features using derivative spectroscopy informed by a trained machine-learning model. 
Several more recent methods also insist on spatial coherence of Gaussian components, such as \textsc{GaussPy+}~\citep{Riener2019}, \textsc{Rohsa}~\citep{Marchal2019}~and \textsc{ScousePy}~\citep{Henshaw2019}. 

Though some of these methods employ elements of Bayesian statistics (i.e. \textsc{ScousePy}~uses the Akaike Information Criterion to resolve discrepancies between spatially neighbouring decompositions) most ultimately rely on a frequentist assessment of the goodness of a particular fit. Therefore these methods are not able to fully incorporate prior information or directly compare the posterior probabilities of competing decomposition models: the main advantages of Bayesian algorithms. 

An example of a fully Bayesian approach to Gaussian decomposition is described in \citet{Allison2012}. The aim of this method is to identify extragalactic H\textsc{i}~absorption in radio data cubes such as those obtained by the Australian Square Kilometre Array Pathfinder (ASKAP). Their method first identifies bright continuum sources in the two on-sky dimensions, then it searches the frequency (equivalently, velocity or redshift) domain for absorption. They utilise \textsc{Multinest}\footnote{https://github.com/farhanferoz/MultiNest}~\citep{Feroz2008,Feroz2009}, a Bayesian Monte Carlo multi-nested sampling algorithm. Though nested sampling methods in general are very powerful and robust \citep[see][for a comprehensive review]{Buchner2021}, they are computationally expensive. The best way to ameliorate this shortcoming is to have very restrictive prior distributions, i.e. to restrict the initial parameter space as much as possible. A primary goal for \amoeba~was for it to be completely automated, so the fine-tuning of the prior distributions required to make a nested sampling approach feasible was deemed prohibitive, and \amoeba~instead applies a Markov Chain Monte Carlo algorithm.

\subsection{A Bayesian approach to the Gaussian Decomposition of OH spectra}
\amoeba~employs a fully Bayesian approach to Gaussian decomposition, allowing the incorporation of prior information and a direct analytical comparison of the posterior probabilities of competing decomposition models. These are related via Bayes' Theorem:
\begin{equation}
{\rm P}(\boldsymbol{\theta}|\boldsymbol{d}, \mathcal{M}) = \frac{{\rm P}(\boldsymbol{d}|\boldsymbol{\theta}, \mathcal{M})~{\rm P}(\boldsymbol{\theta}|\mathcal{M})}{{\rm P}(\boldsymbol{d}|\mathcal{M})}.
\label{Bayes}
\end{equation}

Bayes' Theorem describes the process by which the prior knowledge of the distribution of probability across the possible parameters $\boldsymbol{\theta}$~of a given model $\mathcal{M}$ -- defined by the \textit{a priori}~probability distribution ${\rm P}(\boldsymbol{\theta}|\mathcal{M})$~-- is modified by the presence of the data $\boldsymbol{d}$, the likelihood of which given the possible parameters of the model is described by the distribution ${\rm P}(\boldsymbol{d}|\boldsymbol{\theta}, \mathcal{M})$. This modification of the prior knowledge into the posterior knowledge is via the `evidence' ${\rm P}(\boldsymbol{d}|\mathcal{M})$~which represents the probability of the data $\boldsymbol{d}$~given the model $\mathcal{M}$~\citep{Kass1995}. The ratio of the evidences of two competing models is then equivalent to the odds in favour of one model over the other \citep{Kass1995}. 

The key novelty of \amoeba~is that it simultaneously fits all four ground-rotational transitions of OH, taking advantage of the dependence of directly observable quantities ($\tau_{\nu}$~and $T_{\rm exp}$) on the underlying distribution of OH molecules across the levels of the ground-rotational state, in contrast to typical methods of identifying features in OH spectra.

Some past approaches to the analysis of OH spectra are summarised as follows:
\begin{itemize}
    \item Gaussian features are identified and fit in the main-line transitions only with independent centroid velocities $v$~and full widths at half-maximum $\Delta v$~ \citep[e.g.][]{Li2018,Nguyen2018,Engelke2018,Rugel2018}.
    \item Gaussian features are identified and fit in all four ground-rotational transitions with independent $v$~and $\Delta v$, but with peak values of $\tau_{\nu}$~or $T_{\rm b}$~linked using a statistical equilibrium code such as RADEX \citep{vanderTak2007} or their own molecular excitation code \citep[e.g.][]{Ebisawa2015,Xu2016}.
    \item Gaussian features are identified and fit in all four ground-rotational transitions with the same $v$~and $\Delta v$, and molecular excitation code is used to predict the peak values of $\tau_{\nu}$~or $T_{\rm b}$~based on local cloud properties such as kinetic temperature, number density and local radiation field \citep{Ebisawa2019}.
\end{itemize}

\amoeba~takes advantage of the reasonable assumption that if an OH-containing cloud lies along the line of sight towards a compact continuum source, in the absence of obscuring instrument noise one would expect to observe a Gaussian shaped feature at the same centroid velocity and with the same FWHM in all four ground-rotational state transitions. It also takes advantage of the fact that the peak values of those features ($T_{\rm exp}$, $\tau_{\rm peak}$) are not independent, but ultimately depend on just 5 parameters: the FWHM of the line profile, and the column density in each of the four levels of the ground-rotational state $N_1-N_4$. Therefore, while a single Gaussian feature in a set of on-off spectra may be described by up to 24 parameters (if all 8 spectra are fit independently), \amoeba~will describe the same feature with 6 independent parameters. 

\section{Amoeba}\label{Sec:Amoeba}
    \amoeba~takes as its primary input a set of on-off spectra at each of the OH ground-rotational state transition frequencies. The posterior probability distribution of successively complex models (i.e. with increasing number of Gaussian components) is then sampled until a point is reached where an additional feature does not sufficiently improve the evidence. A model $\mathcal{M}_N$~has $N$~components with the $i$th component having parameters $\boldsymbol{\theta}_{i}=[v_i,$ ~${\rm log}_{10}\Delta v_i,~{\rm log}_{10}N_{1\,i},$ ~$T_{\rm ex}^{-1}(1612)_i,$ ~$T_{\rm ex}^{-1}(1665)_i,$ ~$T_{\rm ex}^{-1}(1667)_i]$, where $v$~is the centroid velocity of the Gaussian feature, $\Delta v$~is its FWHM and $N_{1}$~is the column density in the lowest hyperfine level. $T_{\rm ex}^{-1}(1720)$~is then not a free parameter, but is determined from Eq. \ref{Texsum}. These log and inverse parameters (i.e. $\log_{10}\Delta v,$ ~$\log_{10}N_1,$ ~$T_{\rm ex}^{-1}$) were chosen rather than $\Delta v$, $N_1$, $N_2$, $N_3$~and $N_4$~as their relationship to observable quantities is more linear. From these quantities, the observable quantities $\tau_{\rm peak}$~and $T_{\rm exp}$ are computed from Eqs. \ref{Tex}, \ref{tau} and \ref{Texp}.

Alternatively, \amoeba~can take as input a set of OH optical depth vs. velocity spectra only (hereafter referred to as `optical depth spectra'). In this case the $i$th component of the model will have parameters $\boldsymbol{\theta}_i =$ ~$[v_i,$ ~${\rm log}_{10}\Delta v_i,$ ~$\tau_{\rm peak~1}(1612)_i,$ ~$\tau_{\rm peak~1}(1665)_i,$ ~$\tau_{\rm peak~1}(1667)_i,$ ~$\tau_{\rm peak~1}(1720)_i]$ ~where $\tau_{\rm peak~i}$~are the Gaussian heights at 1612, 1665, 1667 and 1720\,MHz. 

\begin{figure}
    \centering
    \includegraphics[trim={12cm 1.5cm 13.5cm 0.5cm}, clip=true,width=.9\linewidth]{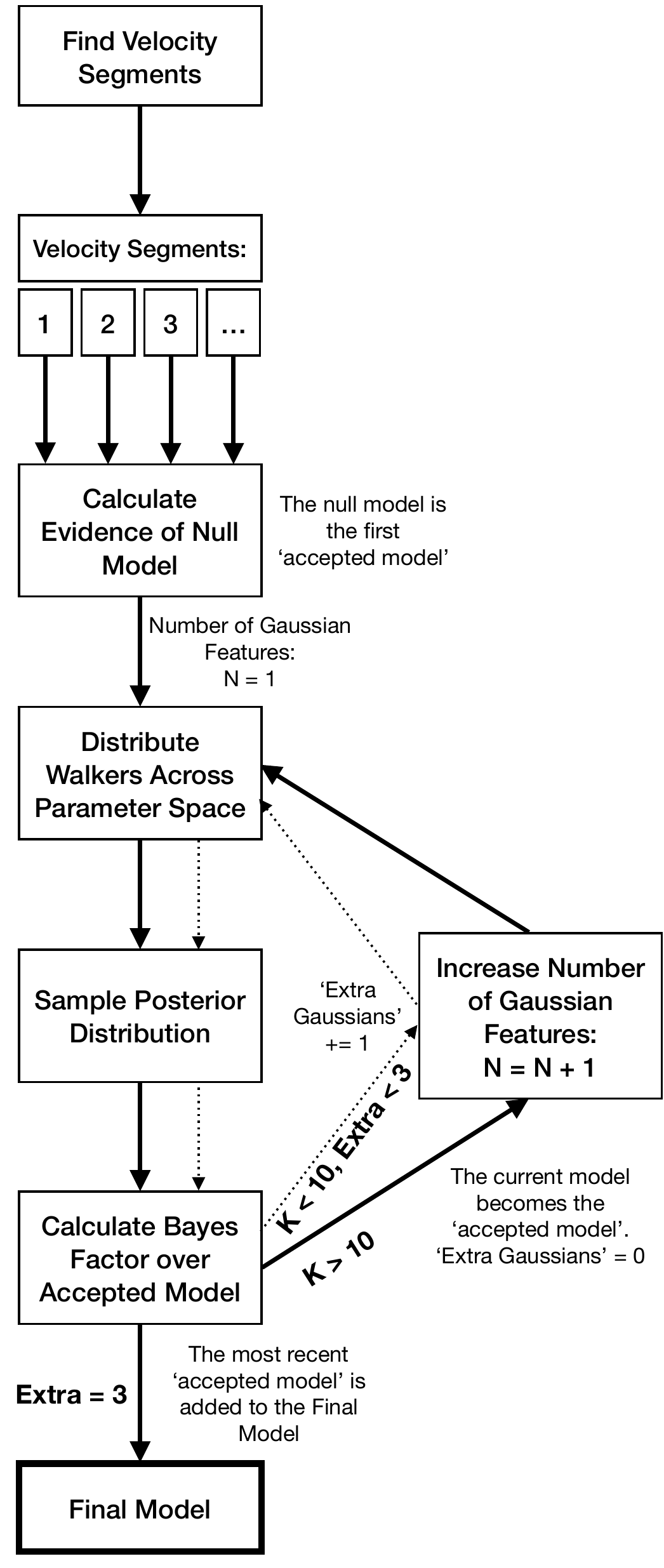}
    \caption{Flowchart illustrating the method by which \amoeba~decomposes hydroxyl spectra into individual Gaussian components.}
    \label{fig:flow}
\end{figure}

\begin{figure*}
    \centering
    \includegraphics[trim={2cm 1cm 2cm 0cm}, clip=true,width=0.9\linewidth]{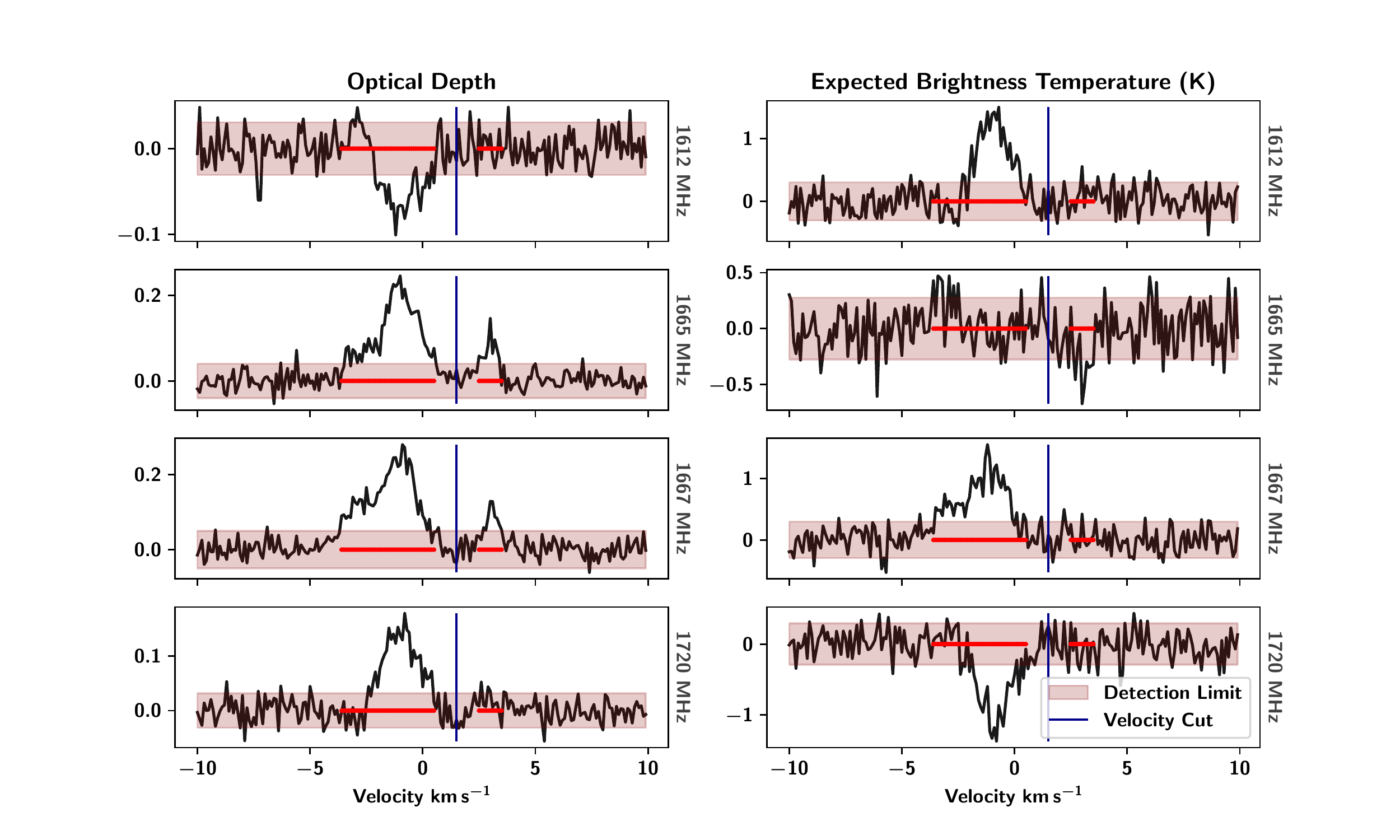}
    \caption{Illustration of the process by which \amoeba~cuts spectra into independent velocity ranges using a set of synthetic data. The user-defined `detection limit' ($1.4~\times$~the rms noise in this example) is indicated by the dark red lines in each plot. The red points indicate all velocities for which at least three consecutive channels in any of the spectra exceeded the detection limit. The velocity cut point is indicated by the vertical blue line.}
    \label{fig:test_vel}
\end{figure*}

The process of Gaussian decomposition is outlined in Fig. \ref{fig:flow}. The first step is to divide the spectra into velocity segments to be processed sequentially. This step serves to reduce computation time and can either be done automatically (via the process explained below) or with velocity ranges provided by the user. Any velocity range extending over a user-defined lower limit of channels (default value is 3 channels) for which the data in all transitions fall below a user-defined `detection limit' is assumed to contain no signal and the spectra are cut at the centre of the velocity range, where the detection limit is a multiple of the rms noise level of each of the provided spectra. This process is illustrated in Fig. \ref{fig:test_vel}. In this example the detection limit is set at $1.4~\times$~the rms noise. Any velocities for which 3 consecutive channels in any of the spectra are above the detection limit are flagged as having significant signal. These are indicated by the red points along the velocity axis in the figure. Any time two ranges of velocities with significant signal are separated by the user-defined lower limit of channels without significant signal, the spectra are all cut at the centre of that range. In the example shown in Fig. \ref{fig:test_vel} the cut point is indicated by the vertical blue line. The resulting velocity segments are assumed to be independent (i.e. they are not expected to include any significant signal from features centred in another velocity segment) and are therefore processed consecutively. The detection limit can be tuned to be quite conservative to avoid the unlikely scenario that a small amount of flux from a Gaussian component may be missed due to the process of cutting the spectra. This process only cuts the spectra into segments, it does not discard any of the data: in the case where no velocity ranges of at least 3 channels (also a user-defined limit) fall above the detection limit, the spectra are not cut but are still carried through to the next step of Gaussian decomposition.

For a given velocity segment, the evidence of the null model is calculated, which is simply the likelihood of a flat model spectrum. The null model is the first `accepted model'. Next is considered the model containing a single Gaussian feature. The joint probability distribution of the model containing a single Gaussian feature is sampled using the \textsc{Python} implementation of Goodman \& Weare's Affine Invariant Markov chain Monte Carlo (MCMC) Ensemble sampler \emcee~\citep{Foreman-Mackey2013}. The evidence of the model is found from a numerical integral of the sampled joint distribution. The smallest volume of parameter space that contributes $68\%$~of the total evidence then approximates the $\pm 1 \sigma$~uncertainties, and the median values in each parameter are reported as the `best' values.

If the evidence of the model containing a single Gaussian is more than a factor of 10 greater than that of the null model \citep[a user-definable limit with a default value based on the standard defined by][]{Jeffreys1961}, the single Gaussian model becomes the current `accepted model' and then the joint probability distribution of the model containing two Gaussian features is sampled, and so on. If the addition of a subsequent Gaussian feature does not result in a factor of ten increase in the evidence, that new model is not accepted, but 3 further models are considered, allowing for the case where the global maximum of the evidence has not yet been found. The number of extra models considered is a user-controlled parameter which can be modified in light of computation cost. In the case where no such models improve the evidence by at least a factor of ten, the current accepted model becomes the final model for the velocity segment, and the process is repeated for the remaining segments. This process is illustrated in the Appendix (Sec. \ref{App:fitting}) with the set of synthetic spectra shown in Fig. \ref{fig:test_vel}.

The evidence is calculated from the relation ${\rm P}(\boldsymbol{d}|\mathcal{M})=\int {\rm P}(\boldsymbol{d}|\boldsymbol{\theta},\,\mathcal{M})\,{\rm P}(\boldsymbol{\theta}|\mathcal{M})\,d\boldsymbol{\theta}$~which follows from the fact that both sides of Eq. \ref{Bayes} integrate to one over all parameter space. Therefore the following two subsections outline how the likelihood distribution ${\rm P}(\boldsymbol{d}|\boldsymbol{\theta},\,\mathcal{M})$~and the \textit{a priori} distribution ${\rm P}(\boldsymbol{\theta}|\mathcal{M})$~are quantified.

\subsection{The likelihood distribution}

As outlined in Sec. \ref{sec:intro}, the joint probability distribution is composed of the likelihood ${\rm P}(\boldsymbol{d}|\boldsymbol{\theta}, \mathcal{M})$~and the prior ${\rm P}(\boldsymbol{\theta}|\mathcal{M})$. The likelihood is defined as 
\begin{equation}\label{likelihood}
    {\rm P}(\boldsymbol{d}|\boldsymbol{\theta}, \mathcal{M})=\frac{1}{\sigma_{\rm noise}^n \sqrt{(2\pi)^n}}\,\exp \left[\frac{-\Sigma_i(d_i-m_i)^2}{2 \sigma_{\rm noise}^2}\right],
\end{equation}
\noindent where the data noise is assumed to be Gaussian, uncorrelated and defined by a standard deviation $\sigma_{\rm noise}$, which is assumed to be constant across a given spectrum but may vary between transitions. The number of data points is given by $n$, the data points are denoted by $d_i$~with corresponding model points $m_i$. \amoeba~can obtain the value of $\sigma_{\rm noise}$~from the input spectra by measuring the standard deviation of several regularly spaced, overlapping segments of each spectrum, then taking the median of these standard deviations. This approach assumes that the majority of these segments are signal-free. If this is not the case, the user may provide appropriate $\sigma_{\rm noise}$~values for the spectra.

\subsection{The a priori distributions}\label{Sec:priors}

\amoeba~assumes a uniform \textit{a priori} probability distribution over the centroid velocity parameter. Additionally, it requires that all centroid velocities fall within the given velocity segment, and in the case of models with more than one Gaussian component, that the centroid velocities are in ascending order. The provision that the centroid velocities be in ascending order is to avoid the consideration of degenerate models that only differ in the ordering of components. Thus the centroid velocity prior distribution for a model $\mathcal{M}_N$~with $N$~Gaussian components takes the form:

\begin{equation}
    {\rm P}(\boldsymbol{\theta}|\mathcal{M}_N)=N!\prod^N_{i=1}\frac{a_i}{v_{\rm max} - v_{\rm min}},
\end{equation}
\begin{equation*}
    \text{where}~a_i=
    \begin{cases}
        1 & \text{if $v_{\rm min} \leq v_i \leq v_{\rm max}$~and $v_i > v_{i-1}$}\\
        0 & \text{otherwise}.
    \end{cases} 
\end{equation*}

\begin{figure*}
    \includegraphics[trim={1.5cm 0cm 2cm 1cm}, clip=true,width=\linewidth]{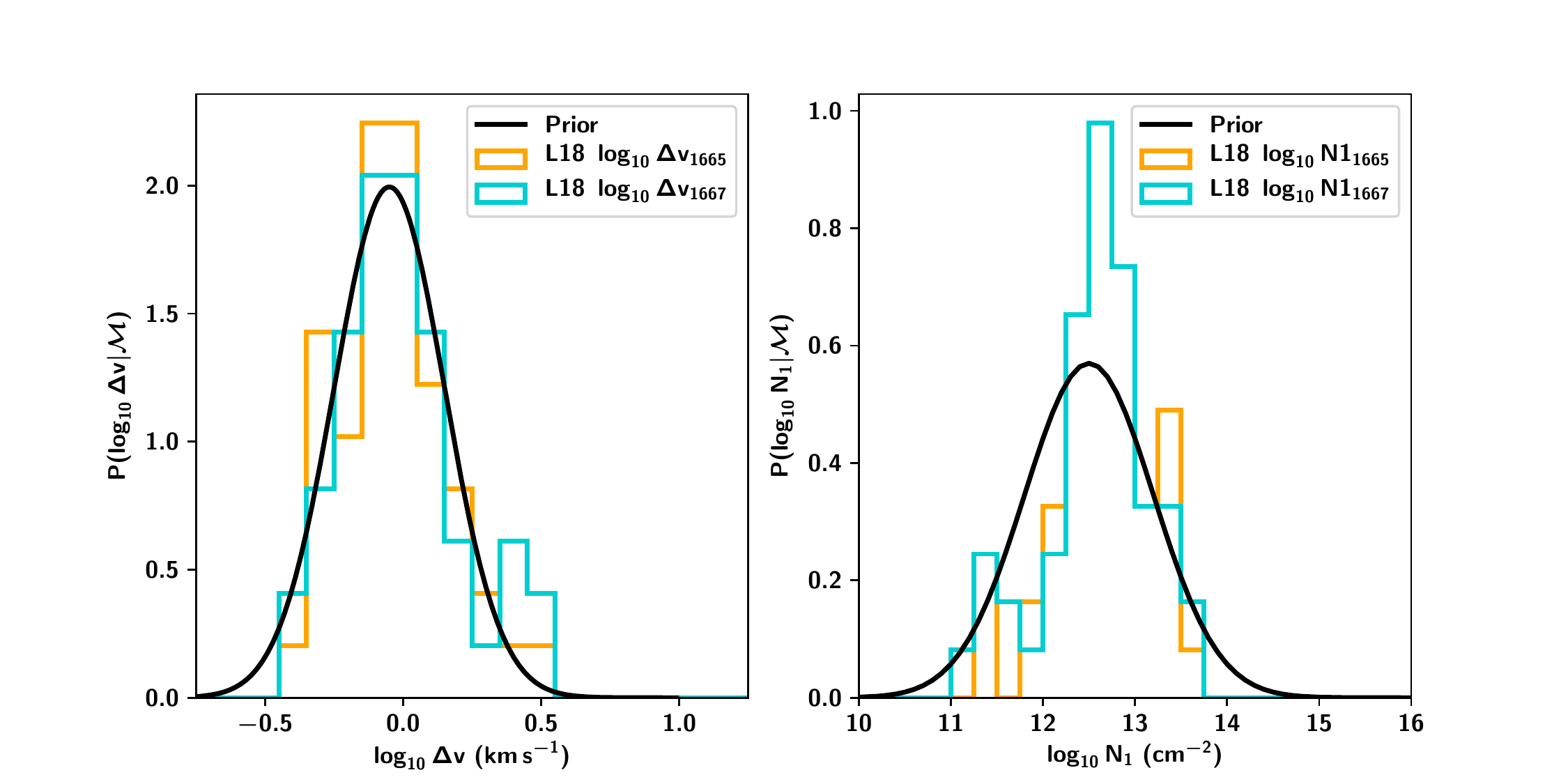}
        \caption{The prior probability distributions for $\log_{10}\,\Delta v$~(left) and $\log_{10}N_1$~(right) plotted over histograms of their values of $\log_{10}N_1$~as determined from 1665\,MHz, and from 1667\,MHz as reported by \citet[][L18]{Li2018}. The prior for $\log_{10}\,\Delta v$~has a default mean of $\mu=-0.5\,$km\,s$^{-1}$~and a default standard deviation of $\sigma=0.2\,$km\,s$^{-1}$, while the prior for $\log_{10}N_1$~has a default mean of $\mu=12.5$\,cm$^{-2}$~and a default standard deviation of $\sigma=0.75$\,cm$^{-2}$.}
    \label{fig:logfwhmlogN1prior}
\end{figure*}

The default \textit{a priori} distributions for log$_{10}\,\Delta v$~and log$_{10}\,N_1$~were loosely based on the values found from the Millennium Survey by \citet[][hereafter L18]{Li2018}, which used the Arecibo telescope to observe the 1665 and 1667\,MHz OH transitions and included sightlines both in and off the Galactic Plane. This survey was chosen for its large sample size, quality of data and derived parameters. The priors and are Gaussian in shape with means $\mu = -0.5\,$km\,s$^{-1}$~and $\mu=12.5$\,cm$^{-2}$, respectively, and standard deviations $\sigma = 0.2\,$km\,s$^{-1}$~and $\sigma=0.75$\,cm$^{-2}$~respectively. These distributions along with values from L18\footnote{L18 reported $N_{\rm OH}$~which is the total OH column density. We converted these to $N_1$~by multiplying their reported value by a factor of 3/16, which assumes a roughly equal population distribution across the sublevels of each of the ground-rotational state levels, which is true in nearly all cases.} are shown in Fig. \ref{fig:logfwhmlogN1prior}. The default \textit{a priori} distribution of log$_{10}\,N_1$~was chosen to be more conservative than the distribution suggested by the results of L18, and is therefore more broad than their results. 

The \textit{a priori} distributions for the remaining parameters of the Gaussian models have defaults informed primarily by the expected distributions given a `reasonable' range of cloud characteristics (outlined in Table \ref{tab:prior_sim}) as determined from non-LTE molecular excitation modelling. This was performed using the models of  \citet{Petzler2020} by choosing random values of each parameter from the distributions noted in Table \ref{tab:prior_sim}, for a total of 10$^6$~models. The default model-based values were cross-checked with the observed distributions from L18.

\begin{table}
    \centering
    \begin{tabular}{lcc}
        \hline
        \hline
        Parameter&Values&Distribution\\
        \hline
        $T_{\rm gas}\,({\rm K})$&10--200&uniform\\
        $T_{\rm dust}({\rm int})\,({\rm K})$&10--50&uniform\\
        $T_{\rm dust}({\rm ext})\,({\rm K})$&10--100&uniform\\
        $\log_{10}{N_{\rm OH}}\,({\rm cm^{-2}})$&$\mu$=13.2, $\sigma$=1&normal\\
        $\log_{10}{\Delta v}\,({\rm km~s^{-1}})$&$\mu$=-0.18, $\sigma$=0.23&normal\\
        $A_v\,({\rm mag})$&0.1, 0.3, 1&choice\\
        $(\log_{10}{n_{\rm H_2}}\,({\rm cm^{-3}}),~X_e)$&(2, -4), (3, -5), (4,-7)&choice\\
        \hline
        \hline
    \end{tabular}
    \caption{Parameters of `typical' ISM clouds input into non-LTE molecular excitation modelling \citep{Petzler2020} used to inform \amoeba's peak optical depth and inverse excitation temperature \textit{a priori}~probability distributions.}
    \label{tab:prior_sim}
\end{table}

\begin{figure*}
    \includegraphics[trim={1.5cm 0cm 2cm 1cm}, clip=true,width=\linewidth]{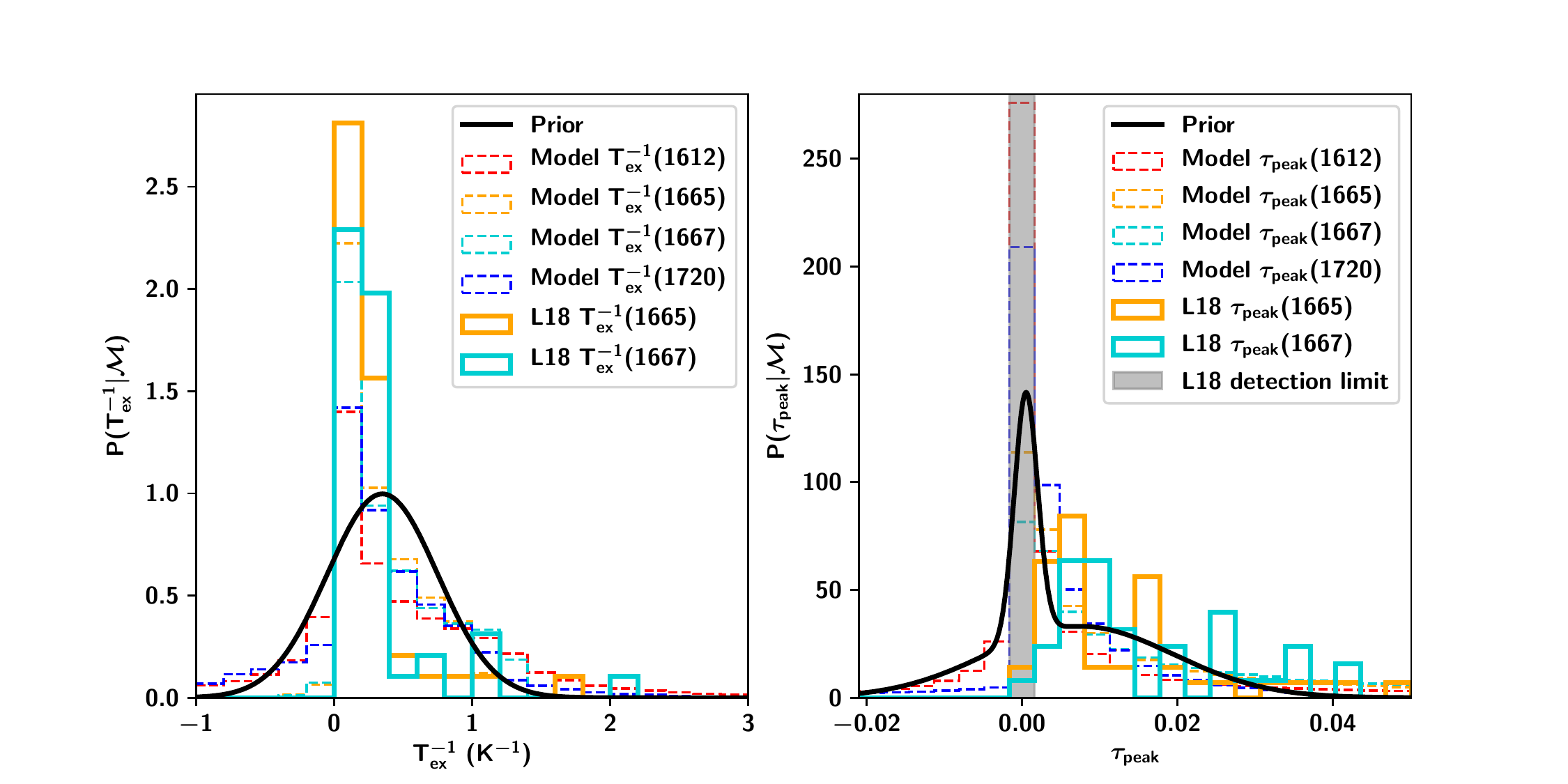}
    \caption{The prior probability distribution for $T_{\rm ex}^{-1}$~(left) and $\tau_{\rm peak}$~(right) plotted over histograms of their values at 1665\,MHz, and at 1667\,MHz as reported by \citet[][L18]{Li2018} and those found from non-LTE molecular excitation modelling as described in the text. The average $1\sigma~\tau_{\rm peak}$~detection limit of L18 is indicated by the grey shaded area. The edges of the histograms were chosen so that this range is a single bin centred at 0. The dashed histograms represent the distribution of $T_{\rm ex}^{-1}$~and $\tau_{\rm peak}$~ obtained from our molecular excitation modelling for a range of `reasonable' ISM cloud parameters (see Table \ref{tab:prior_sim}). The prior for $T_{\rm ex}^{-1}$~has a default mean of $\mu= 0.35\,{\rm K}^{-1}$~and a default standard deviation of $\sigma=0.4\,{\rm K}^{-1}$. The prior for $\tau_{\rm peak}$~has two Gaussian components, the first with mean $\mu=0.01$, standard deviation $\sigma=0.01$~with a weight of 0.6, and the second with $\mu=0.0025$, $\sigma=0.0015$~and a weight of 0.4.}
    \label{fig:invTextauprior}
\end{figure*}

The default \textit{a priori}~distributions for both $T_{\rm ex}^{-1}$~and $\tau_{\rm peak}$~are guided by the distributions suggested by our non-LTE modelling and the results of L18, which are shown in Fig. \ref{fig:invTextauprior}.

Given that \amoeba~is intended as an automated algorithm for large datasets, we chose default \textit{a priori}~distributions somewhat simpler than the distributions suggested by our modelling and the results of L18, primarily to avoid unacceptable biases from over-tuned priors. This aim led us to chose a Gaussian-shaped prior for $T_{\rm ex}^{-1}$~with mean $\mu=0.35\,{\rm K}^{-1}$~and standard deviation $\sigma=0.4\,{\rm K}^{-1}$. The distribution for $\tau_{\rm peak}$~seen in our modelling and L18 had a narrower peak and wider, less symmetric wings than that seen in $T_{\rm ex}^{-1}$, so we chose a double-Gaussian prior for $\tau_{\rm peak}$, with means $\mu=0.01,\,0.0025$, standard deviations $\sigma=0.01,\,0.0015$, and relative weights of 0.6 and 0.4, respectively. These default \textit{a priori}~distributions are shown in Fig. \ref{fig:invTextauprior}. 

Though care was taken when choosing these default priors, we urge users of \amoeba~to assess whether they are appropriate for their intended use case. For instance, for data towards the Central Molecular Zone or even some parts of the Inner Galaxy, one might expect much larger FWHM or higher optical depths. In that case the default priors will introduce significant bias and these parameters will tend to be under-estimated. We illustrate this in Section \ref{sec:Test1} by showing that for synthetic spectra generated with parameters that have a very low prior probability the fitted parameters will show a small but systematic shift (clearly seen for optical depth in Fig. \ref{fig:Test1_tau}) towards the values preferred by the priors. The user can avoid this by choosing a more appropriate or wider prior distribution for these parameters.

\section{Performance and Discussion}\label{Sec:Discussion}
    \amoeba's performance was assessed in a series of tests which examined the validity of its core Bayesian algorithm (see Sec. \ref{sec:sbc}), its ability to recover `correct' parameters (see Sec. \ref{sec:Test1}), its sensitivity to noise (see Sec. \ref{sec:Test2}), and its ability to resolve closely blended features (see Sec. \ref{sec:Test3}). These tests were performed on synthetic data with a range of parameters outlined in Table \ref{tab:test_params}. In all but the last test the synthetic data were generated with a single Gaussian feature, and in the last test two identical Gaussian features were present. Noise was added to these spectra and they were then decomposed. All tests were performed with both on-off spectra and optical depth spectra only. \amoeba~performed well in these tests, as is discussed further in this section.

\begin{table}
    \centering
    \begin{tabular}{cc}
        \hline
        \hline
        Parameter&Range\\
        \hline
        $v~({\rm km\,s}^{-1})$&-8 -- 8\\
        $\Delta v~({\rm km\,s}^{-1})$&0.3 -- 5\\
        $\log_{10}N_1~({\rm cm}^{-2})$&11 -- 15\\
        $T_{\rm ex}~({\rm K})$&-20 -- 20\\
        $\tau_{\rm peak}$&-0.05 -- 0.05\\
        \hline
        \hline
    \end{tabular}
    \caption{The range of parameters used to generate synthetic spectra when testing the ability of \amoeba~to recover `correct' parameters. All values were chosen from a uniform distribution in the stated range.}
    \label{tab:test_params}
\end{table}

\subsection{Validity of Bayesian algorithm}\label{sec:sbc}
The validity of \amoeba's Bayesian inference algorithm was tested using `simulation-based calibration' (SBC) following the method described by \citet{Talts2018}. Though Bayesian analysis is itself straightforward in its application of Bayes Theorem (Eq. \ref{Bayes}), algorithms built to handle complex models and data sets can easily fail to reach satisfactory conclusions if the algorithm lacks certain self-consistencies. For example, the choice of a given parameter of a model -- say the FWHM $\Delta v$~of a Gaussian-shaped feature -- may introduce unintended biases to the algorithm, whereas a different choice -- i.e. its logarithm $\log_{10}\Delta v$~-- may eliminate this bias. SBC aims to identify and diagnose these issues through a straightforward test, the results of which are described in Section \ref{sec:sbc}.

\begin{figure*}
\centering
    \includegraphics[trim={1cm 1cm 2cm 3cm}, clip=true, width=.8\linewidth]{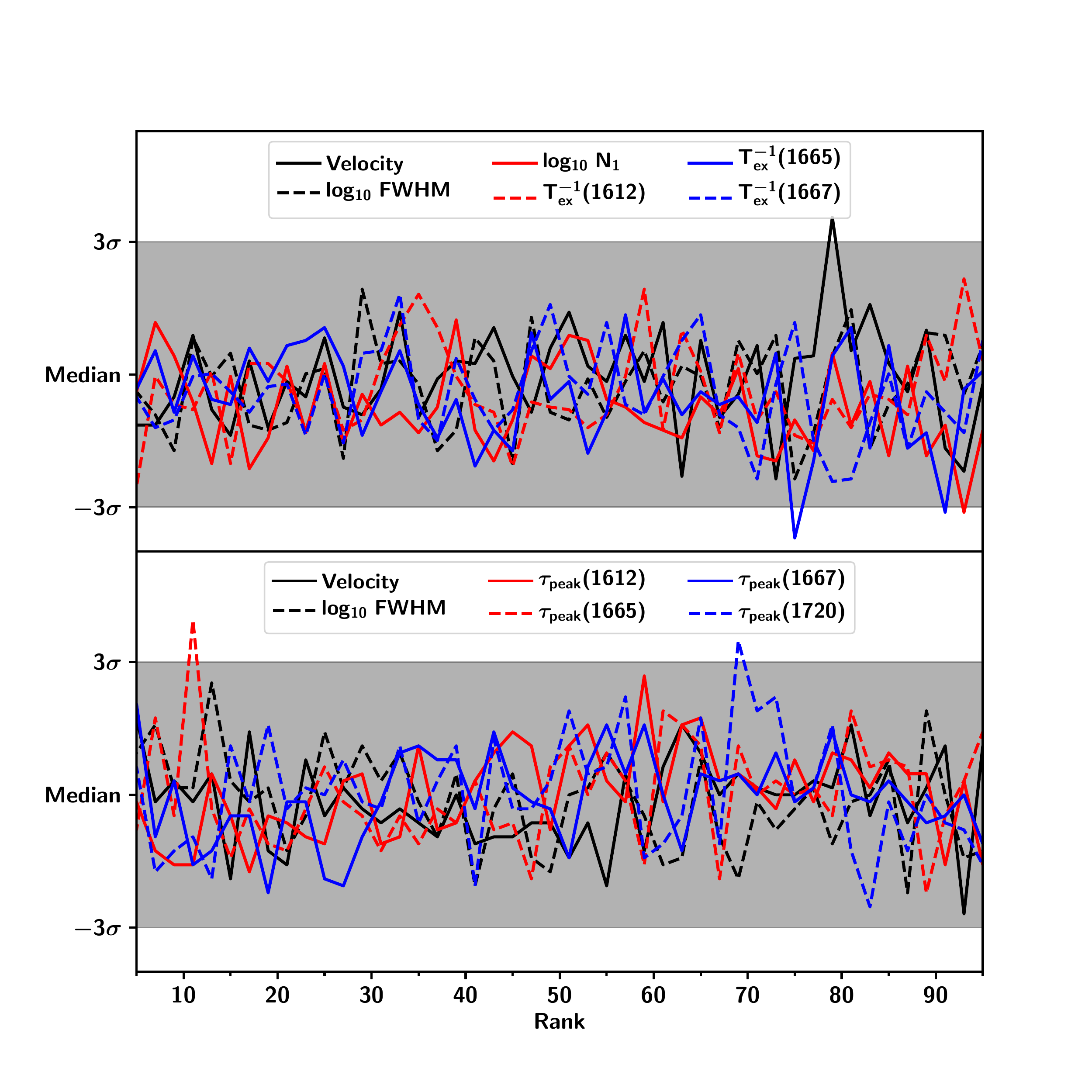}\\ 
    \caption{Histograms of the ranks obtained through simulation-based calibration of \amoeba~for synthesised data with both on- and off-source observations (top) and for synthesised data with only optical depth spectra (bottom). The peaks of each histogram bin are plotted as points on a line for visual clarity. Each line represents an individual model parameter as described in the text. A well-calibrated Bayesian algorithm is expected to have a uniform distribution across all ranks. The shaded region indicates the area expected to contain 99.7\% of the data points if the distribution was uniform.}
    \label{fig:sbc_hist}
\end{figure*}

In this test, a set of parameters representing a single Gaussian-shaped profile feature:
\begin{align*}
\tilde{\boldsymbol{\theta}}=[&v,~\log_{10}\Delta v,~\log_{10}N1,~T_{\rm ex}^{-1}(1612),\\&T_{\rm ex}^{-1}(1665),~T_{\rm ex}^{-1}(1667)]
\end{align*}
\noindent for on-off spectra, or 
\begin{align*}
\tilde{\boldsymbol{\theta}}=[&v,~\log_{10}\Delta v,~\tau_{\rm peak}(1612),\\&\tau_{\rm peak}(1667),~\tau_{\rm peak}(1720)]
\end{align*}
\noindent for optical depth only, were drawn from the prior distribution P($\boldsymbol{\theta}|\mathcal{M}$) and a sample data set was constructed by drawing from the likelihood distribution P($\boldsymbol{d}|\boldsymbol{\theta},\mathcal{M}$). \amoeba~was then used to sample the posterior probability distribution, returning the converged Markov chains. If \amoeba's Bayesian inference algorithm is valid, the originally drawn set of parameters $\tilde{\boldsymbol{\theta}}$~will have an equal probability of falling at any percentile rank within the converged Markov chains \citep[see][for a detailed proof]{Talts2018}. Therefore, if this process of selecting a set of parameters from the prior distribution, simulating data by drawing from the likelihood distribution and then sampling the posterior probability distribution is repeated many times, the distribution of the percentile ranks of those drawn parameters within the converged Markov chains will show a flat distribution.
The distribution of these percentile ranks for 2000 trials are shown in Fig. \ref{fig:sbc_hist}. 

Fig \ref{fig:sbc_hist} shows ranks from 5 to 95, as the edge bins will tend to be over-populated due to auto-correlation in the sample set. This auto-correlation is found in all samples drawn using MCMC and indicates limitations in the simulation-based calibration test rather than the Bayesian algorithm \citep{Talts2018}. The shaded regions of Fig. \ref{fig:sbc_hist} indicate the area that should contain 99.7\% of the data points shown if their distribution was truly uniform: only 1 point should lie outside the shaded region. For the method using on-off spectra, 4 points fell outside this shaded region, while for optical depth spectra, 2 points fell outside. Overall 99\% of the points fall within the required range, implying no significant biases in our Bayesian algorithm.

\subsection{Ability to recover `correct' parameters}\label{sec:Test1}
\begin{figure*}
    \centering
    \includegraphics[trim={1cm 1.5cm 1cm 3.1cm}, clip=true, width=.9\linewidth]{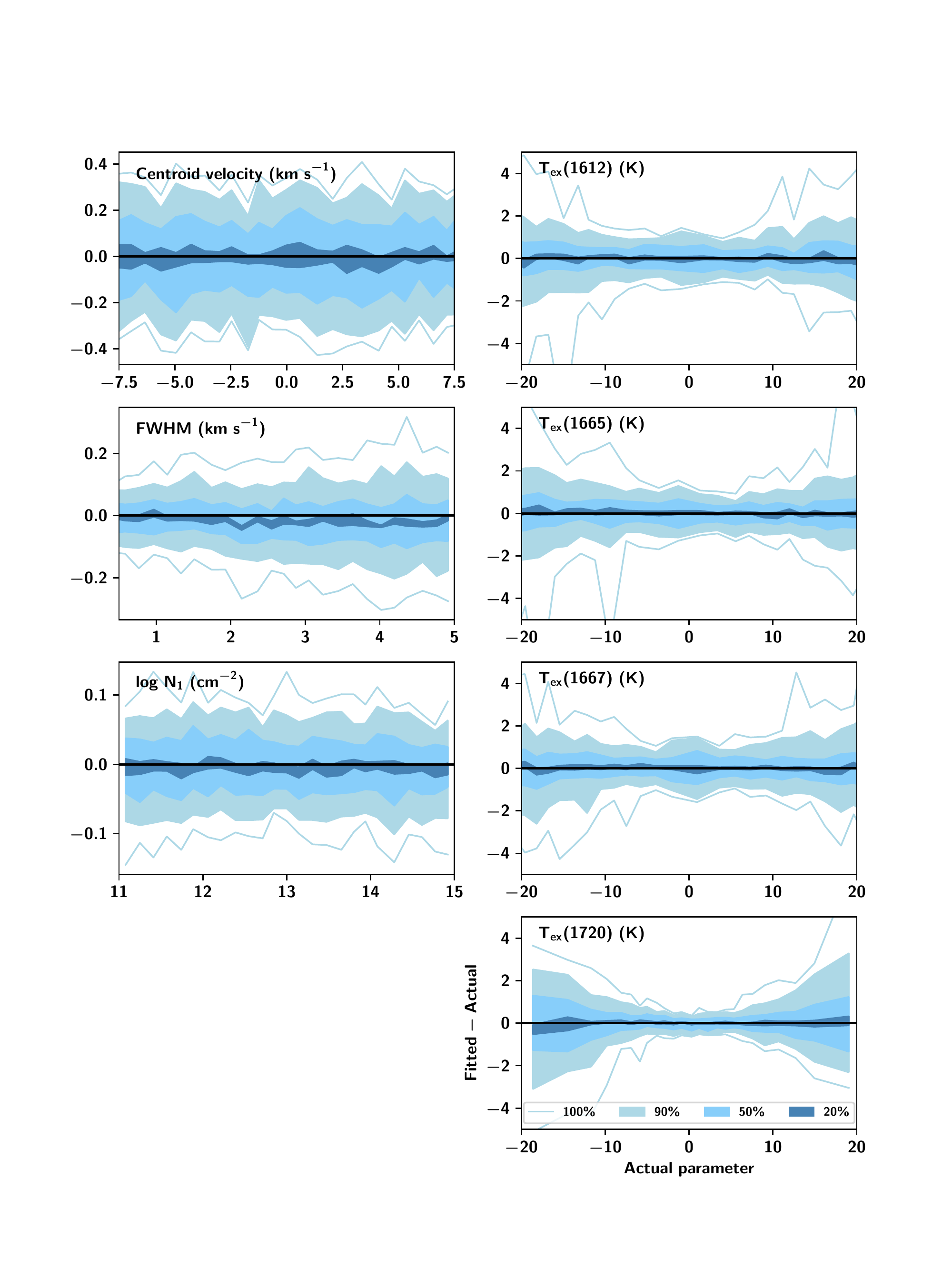}
    \caption{Difference between the final parameters recovered by \amoeba~and the `actual' parameters for `on-off' spectra synthesised with a signal-to-noise ratio of 5 and a channel width of 0.1 km\,s$^{-1}$. The shaded regions represent the inner 90, 50 and 20\% range of all trials, with each bin on the x-axis containing 100 trials. The region enclosing all trial results (the 100\% range) is indicated by the pale blue line. For the tests shown there were no false positives or false negatives.}
    \label{fig:Test1}
\end{figure*}

\begin{figure*}
    \centering
    \includegraphics[trim={0.5cm 1cm 1.5cm 2cm}, clip=true, width=.9\linewidth]{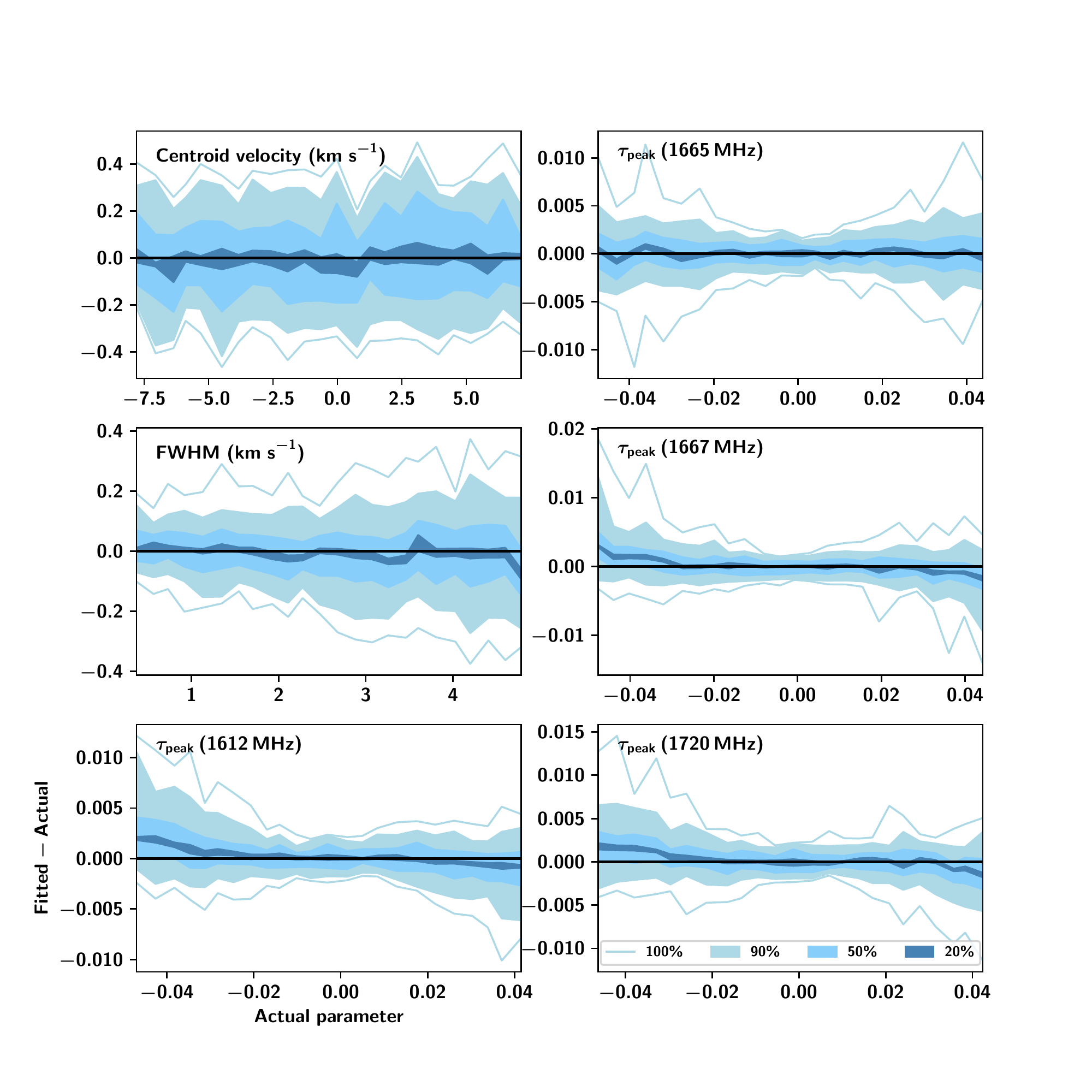}
    \caption{Same as Fig. \ref{fig:Test1} but for synthetic optical depth spectra.  For the tests shown there was a false positive rate of 0.04\%~and there were no false negatives.}
    \label{fig:Test1_tau}
\end{figure*}

The purpose of this test was to quantify the ability of \amoeba~to recover true parameters from data by fitting synthetic data with known parameters. It also served to test the sensitivity of the parameter recovery rate to the \textit{a priori} distributions. In this test synthetic spectra were generated with a range of parameters (see Table \ref{tab:test_params}) with noise added such that each spectrum had a signal-to-noise ratio of 5. This signal-to-noise ratio is intended to represent a reasonably strong signal given the weakness of most diffuse OH features \citep[e.g.][]{Li2018,Rugel2018}. \amoeba~was then used to recover the parameters. From 5000 trials none produced a false negative result, and only 7 of the on-off spectra returned a false positive result. Figs. \ref{fig:Test1}~and \ref{fig:Test1_tau}~then show the difference between the final fitted parameters returned by \amoeba~to the original parameters for the remaining spectra. The shaded regions in these figures show the range covered by the inner 20, 50 and 90\% of tests within each bin of 100 samples; the outer line shows the full 100\% of samples.

Testing over a FWHM range of 0.3--5.0\,km\,s$^{-1}$, we find that 50\% of the synthetic on-off spectra and optical depth spectra the centroid velocity was within 1 channel (0.1\,km\,s$^{-1}$) of the correct location, and in both cases was within 2 channels of the correct location for 90\% of the spectra. We note however that the 68\% credibility intervals reported by \amoeba~for the centroid velocity parameter tends to not reflect this spread: the typical 68\% credibility interval covers $\approx 0.01\,$km\,s$^{-1}$. This is a reflection of our treatment of the data as an incomplete sampling of a continuous Gaussian distribution, rather than considering the individual channels as bins of that continuous function. The overall effect of this simplification is that the credibility interval for centroid velocity (and FWHM, discussed further below) will under-estimate the true uncertainty. 
We therefore suggest users of \amoeba~adopt a more conservative estimate of the uncertainty of these parameters: $\approx \pm  1$~channel width is likely a more accurate reflection of the effective uncertainty at this signal-to-noise level.

For values of $\sim$1\,km\,s$^{-1}$ (typical of the L18 sample and hence well matched to our chosen \textit{a priori} distribution), the FWHM recovered was within one channel width of the correct value for 90\% of both the synthetic on-off spectra and the optical depth spectra. For larger linewidths the accuracy of the recovered value is reduced due to the influence of the \textit{a priori} distribution. We therefore advise if wide features are expected (i.e. $\Delta v>4$\,km\,s$^{-1}$) that the \textit{a priori} distribution be adjusted accordingly. 

Testing over a range of $-$20 to $+$20\,K, we find that the recovered excitation temperatures were within $\approx 10\%$~of the actual values down to an accuracy of $\pm 1\,{\rm K}$~for 90\% of on-off spectra. We note that Fig. \ref{fig:Test1} implies a reduction in accuracy for $T_{\rm ex}\,(1720)$~when $|T_{\rm ex}\,(1720)|>5\,$K. This is a reflection of the fact that $T_{\rm ex}^{-1}\,(1720)$~is not a free parameter in our model, and therefore the values of $T_{\rm ex}\,(1720)$~did not have a uniform distribution in this test.

Testing over a $\log N_1$~range of 11 to 15 (cm$^{-2}$), the recovered column density was within $\sim 10$\% of the correct value for 90\% of the synthetic on-off spectra. This is a very encouraging result when compared to the uncertainties found by L18 which tended towards 100\%, or \citet{Nguyen2018} which tended towards 40\%. 

In optical depth, for a test range of $-$0.05 to $+$0.05, 60\% of recovered values were within within $\pm 10\%$~of the actual values. In the case of optical depths at 1612, 1667 and 1720\,MHz, the fitted values tend towards a lower magnitude than the actual values, as a result of the narrow optical depth \prior~distribution. We therefore advise that in cases where higher optical depths are expected (i.e. $|\tau| > 0.05$) that \amoeba's prior distributions be modified accordingly.

\subsection{Effect of signal-to-noise ratio}\label{sec:Test2}
The purpose of this test was to quantify the signal-to-noise ratio over which \amoeba~could be relied upon to recover low-signal features. 
In this test, spectra with varying noise levels and feature parameters were generated and \amoeba~was then used to recover those parameters. The aim of this test was to determine the relationship between signal-to-noise ratio and \amoeba's ability to produce `good-quality' fits to the data. 

In order to describe the quality of the fits returned by \amoeba, we define four broad categories that we use in this and the next test: well fit, poorly fit, false negative and false positive. The fits were placed in these categories by the following method. Any fits where fewer than the correct number of features were identified were immediately categorised as false negatives. Similarly, any fits where more than the correct number of features were identified were immediately categorised as false positives. Then all fits with the correct number of features were checked to see if all the true parameters fell within the 99.95\% credibility interval, as approximated by three times the 68\% credibility interval. We found that in general the posterior probability distribution was Gaussian, implying that three times the 68\% credibility interval is a good approximation of the 99.95\% credibility interval. (Typical corner plots are shown in Figs. \ref{fig:corner_Texp} and \ref{fig:corner_tau} in the Appendix). If all parameters of a fit were within the 99.95\% credibility interval, it was categorised as well fit. If at least one but not all were in the 99.95\% credibility interval it was categorised as poorly fit, otherwise it was categorised as a false positive. False positives identified in this way would also technically be false negatives, but we felt that false positives were the more concerning behaviour, and therefore categorised them as such.

\begin{figure*}
    \centering
    \begin{tabular}{c|c}
        \includegraphics[trim={1cm 0.5cm 1cm 1cm}, clip=true, width=0.6\linewidth]{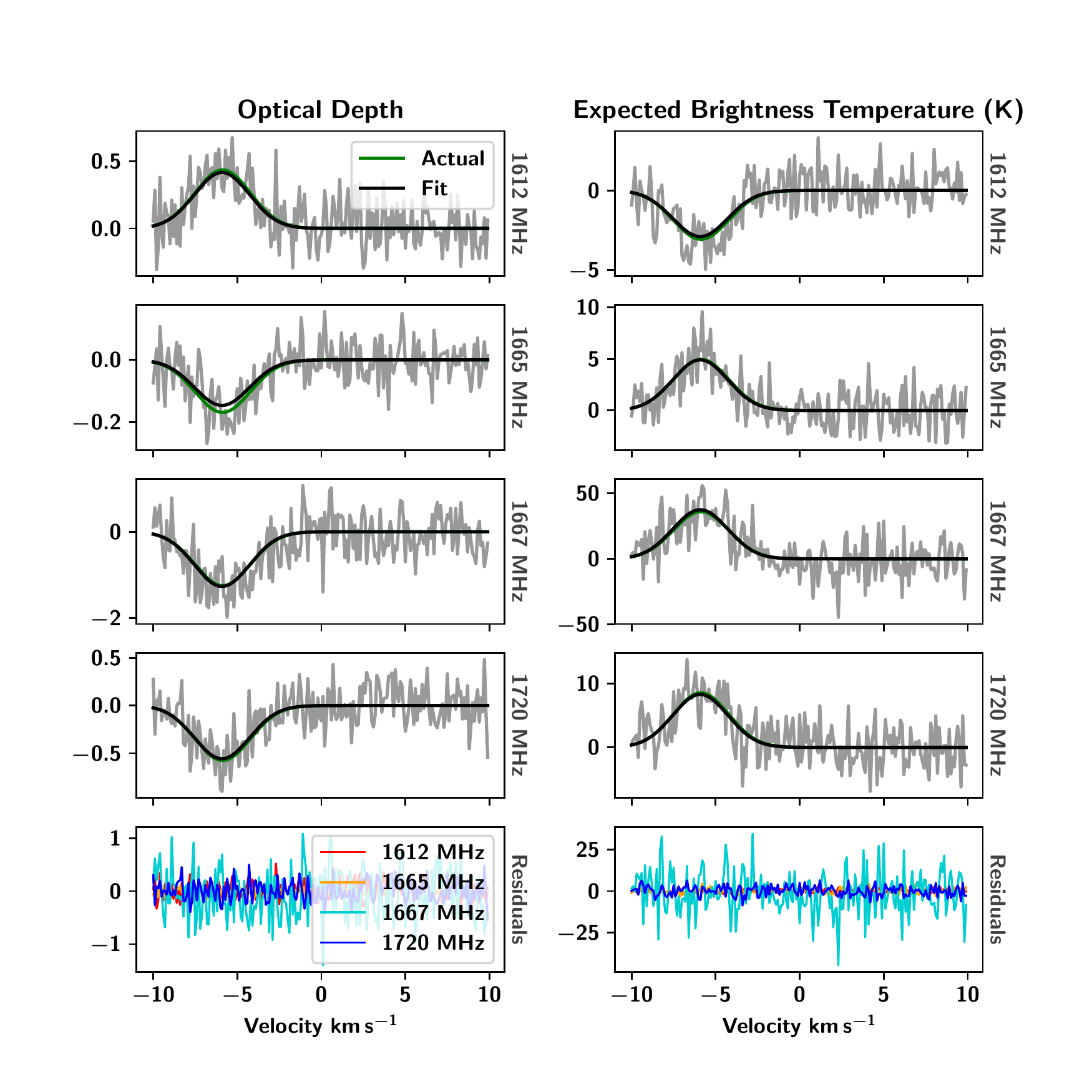}&
        \includegraphics[trim={.5cm 0.5cm 9cm 1cm}, clip=true, width=0.315\linewidth]{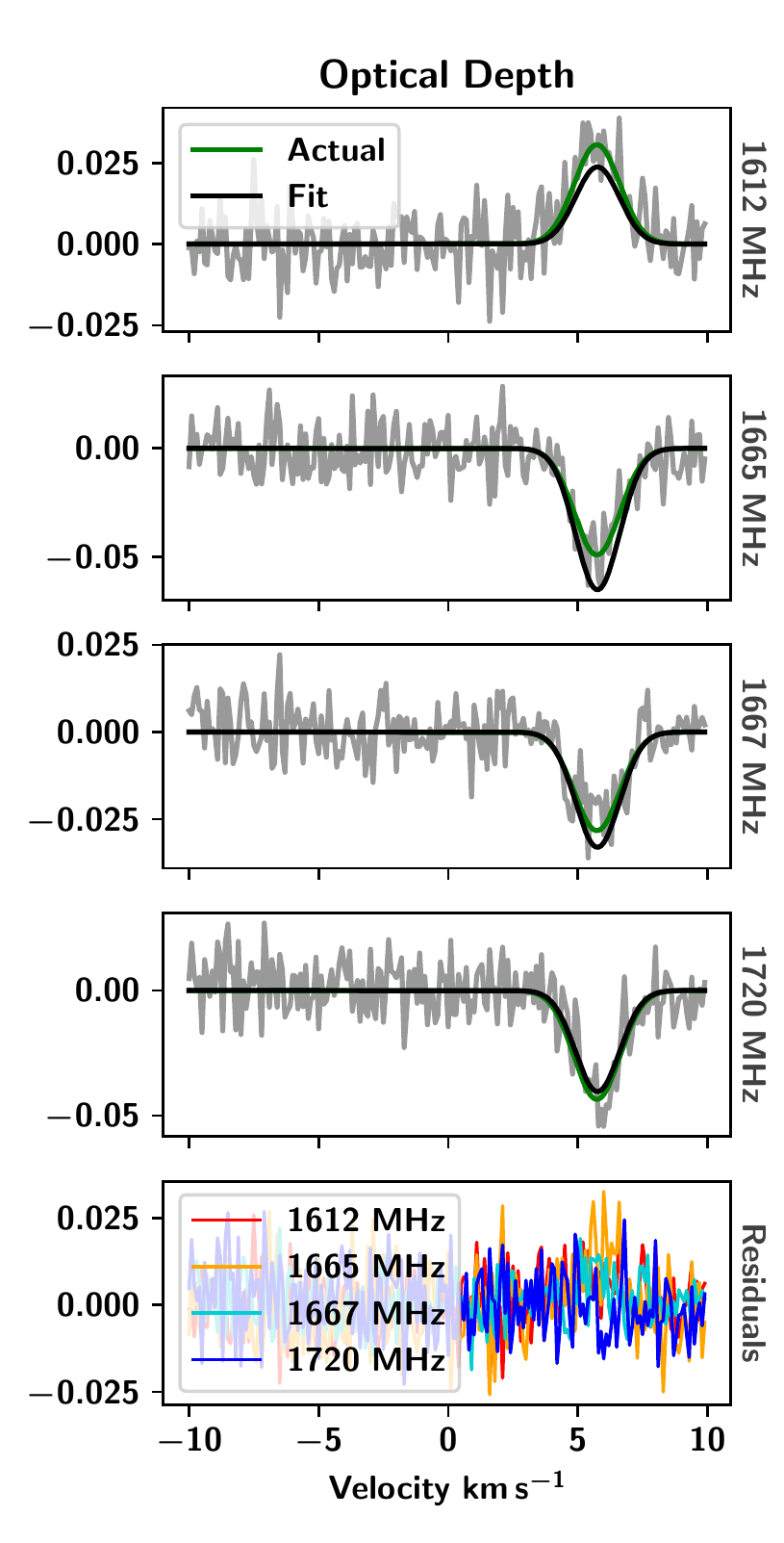}\\
    \end{tabular}
    \caption{Typical examples of synthetic spectra `poorly fit' by \amoeba~while testing the effect of signal-to-noise ratio. The first two columns show a single set of synthetic on-off spectra with optical depth at the left and expected brightness temperature at the right. The third column shows a different set of synthetic optical depth spectra only. The top row of each column show the 1612\,MHz synthetic data (grey), the noiseless `Actual' feature (green) and the fit returned by \amoeba~(black). The second, third and fourth rows then show the 1665, 1667 and 1720\,MHz transitions, respectively. The bottom row shows the residuals of each transition. For both the on-off spectra (left) and the optical depth spectra (right) the actual values of the spectra parameters did not fall within the 99.95\% credibility interval of the sampled posterior probability distribution.}
    \label{fig:PF_2}
\end{figure*}

This seemingly wide range in the credibility interval was chosen upon examination of a selection of the fits which showed that using the same steps with a narrower credibility interval threshold would categorise a significant number of visually good fits as poor or false positives. Even with our broad credibility interval threshold most fits categorised as `poor' are not obviously `wrong' to the eye (see an example of a `poor' fit in Fig. \ref{fig:PF_2}). This behaviour was mostly a reflection of the narrow credibility intervals often returned for some parameters, notably the FWHM. Indeed, as can be seen from the results of the previous test outlined in Section \ref{sec:Test1}, although there is a spread in the returned parameter values compared to the actual values, there are few if any extreme outliers at a signal-to-noise level of 5 (see Figs. \ref{fig:Test1} and \ref{fig:Test1_tau}). i.e. even `poor' fits in the present test are unlikely to correspond to pathological deviations in the true physical parameters. 

\begin{figure*}
    \centering
    \includegraphics[trim={0cm 0cm .5cm .5cm}, clip=true, width=0.475\linewidth]{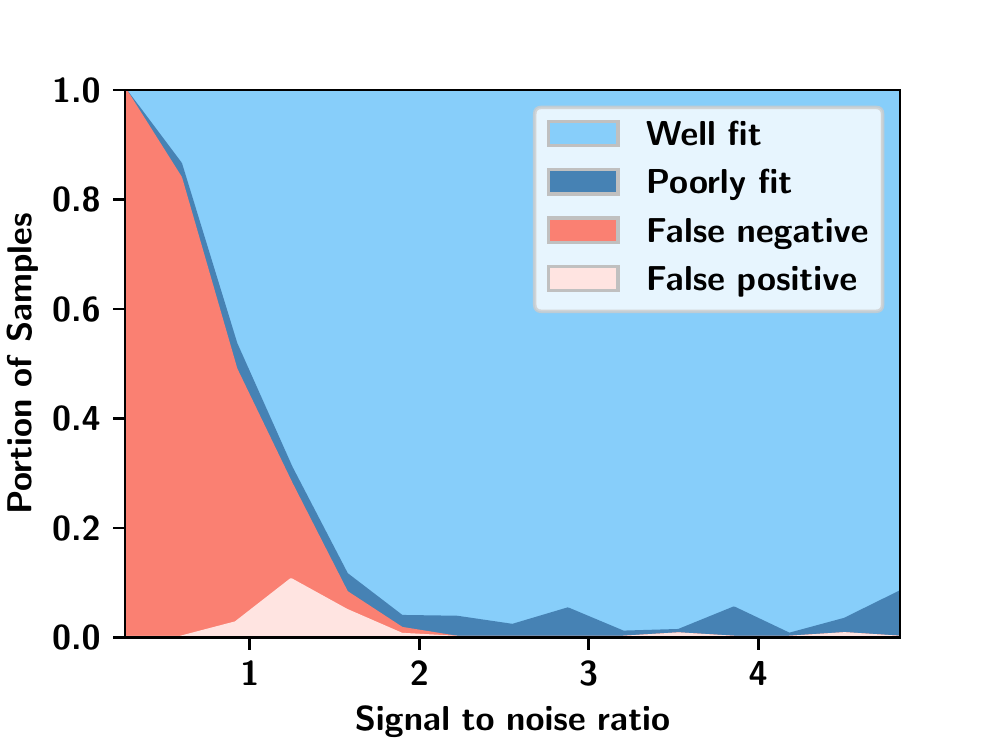}\includegraphics[trim={0cm 0cm .5cm .5cm}, clip=true, width=0.475\linewidth]{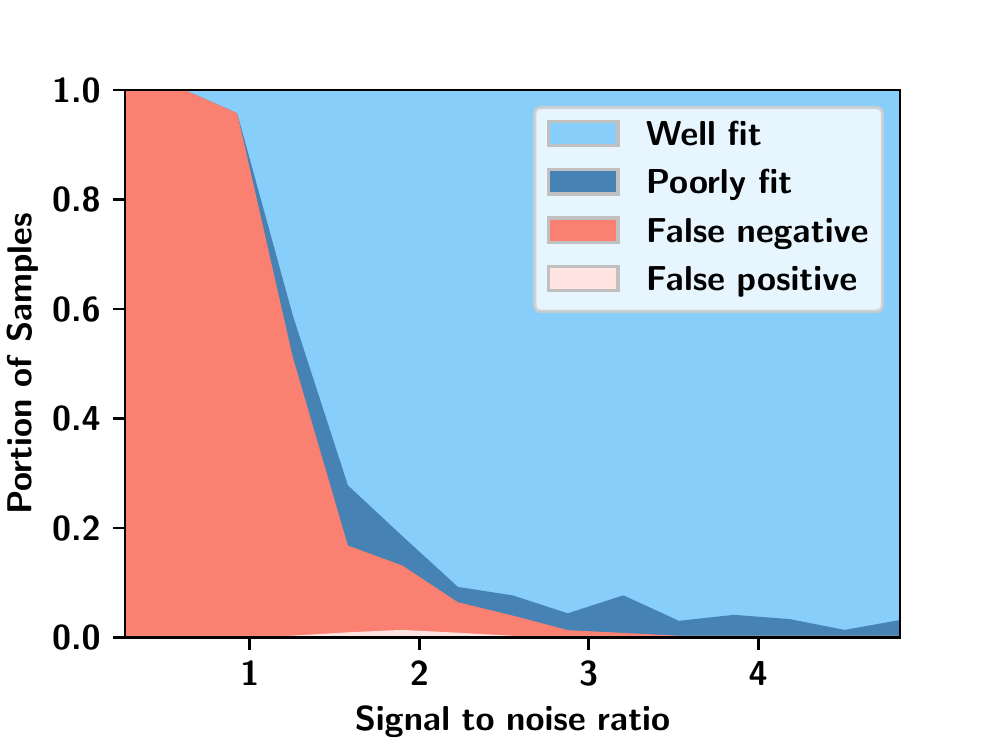}\\
    \caption{Fraction of synthetic `on-off' spectra (left) and optical depth spectra (right) for which the 99.95\% credibility interval returned by \amoeba~contained all the correct parameters~(`Well fit'), contained at least one but not all of the correct parameters~(`Poorly fit'), did not contain any of the correct parameters or where \amoeba~failed to identify any Gaussian features (`False negative'), or where \amoeba~recovered more than one feature (`False positive'), plotted against the signal-to-noise ratio of all synthesised spectra.}
    \label{fig:Test2}
\end{figure*}

The results of this test for on-off synthetic spectra as well as optical depth-only synthetic spectra are illustrated in Fig. \ref{fig:Test2}. As expected, at signal-to-noise ratios below 1 the majority of spectra resulted in a false negative result: the noise was too high to justify the acceptance of a feature. In both cases this false negative rate dropped quickly with increased signal-to-noise ratio, and the number of `well fit' spectra reached 90\% at a signal-to-noise ratio of 2~for the synthetic on-off spectra and at 3~for the synthetic optical depth spectra. Both had a similar rate of `poorly fit' results. The ability of \amoeba~ to produce reliable fits to the majority of on-off spectra even at the 2$\sigma$ level is a reflection of the power of fitting all four transitions simultaneously, and represents a significant improvement over traditional decoupled fitting.

\subsection{Effect of feature separation}\label{sec:Test3}
\begin{figure*}
    \centering
    \begin{tabular}{ll}
    \includegraphics[trim={0cm 0cm .5cm .5cm}, clip=true, width=0.475\linewidth]{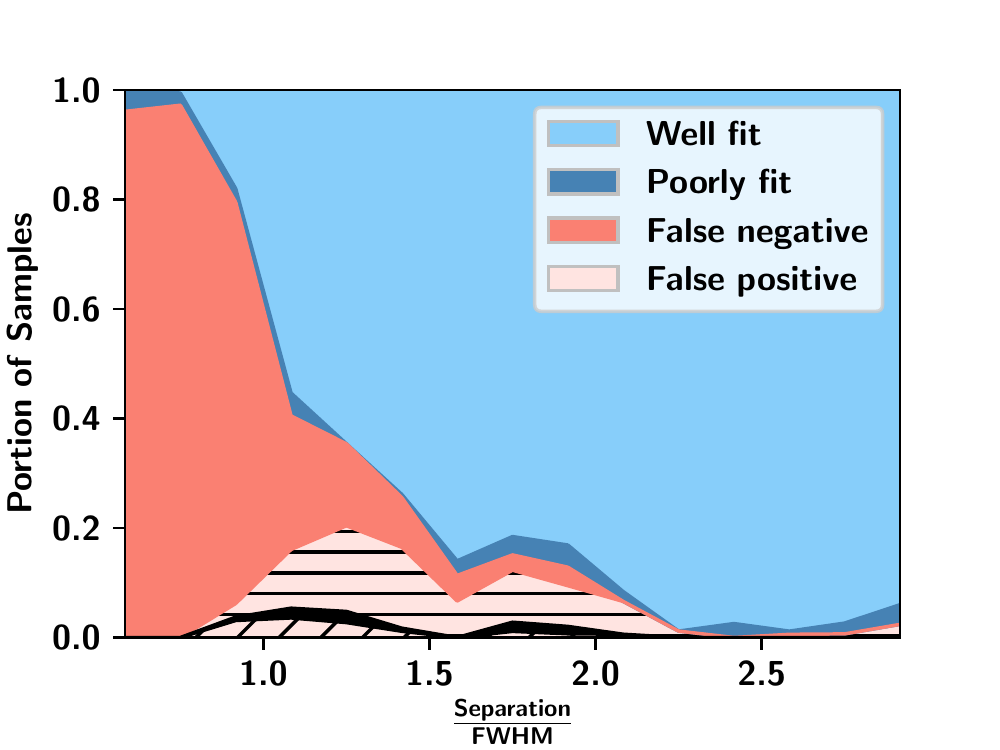}&\includegraphics[trim={0cm 0cm .5cm .5cm}, clip=true, width=0.475\linewidth]{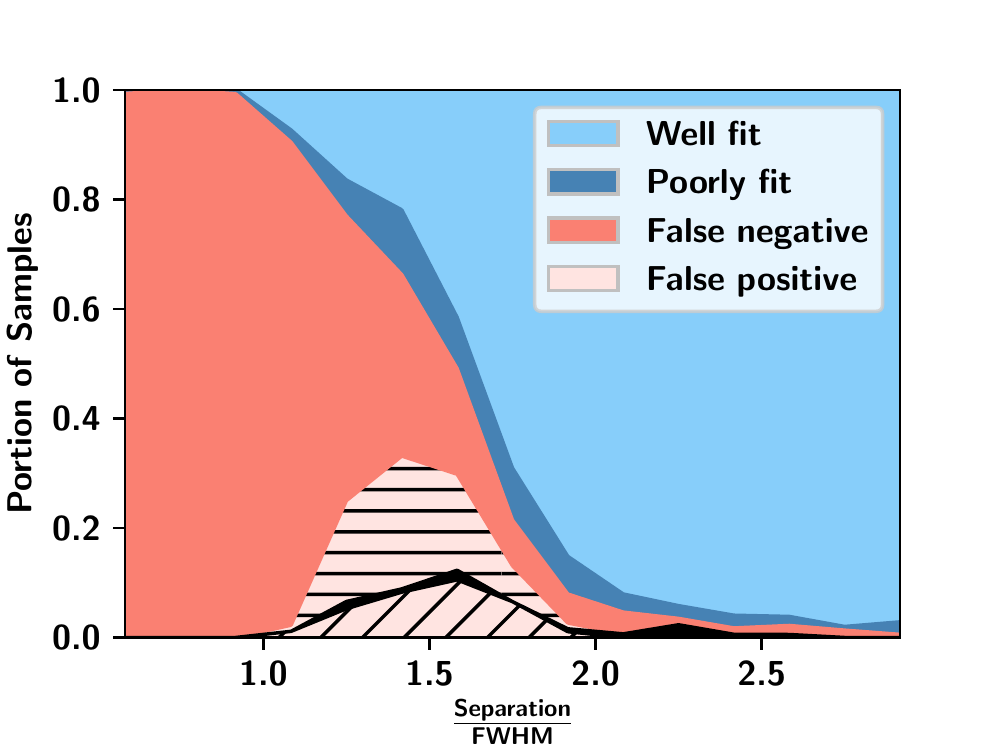}\\
    \end{tabular}
    \caption{Fraction of synthetic `on-off' spectra (left) and optical depth spectra (right) for which \amoeba~recovered all parameters within the 99.95\%~credibility interval~(`Well fit'), recovered the correct number of features but where not all (but at least 1) of the parameters were within the 99.95\%~credibility interval~(`Poorly fit'), failed to resolve the two features present (`False negative'), and recovered more than two features, or features for which none of the parameters fell within the 99.95\%~credibility interval (`False positive'), plotted against the separation of the two features as a multiple of the full width at half-maximum. The false positives are then divided into those where neither of the two features were recovered (diagonal hatch), where one was well fit (solid black) and where both were well fit (horizontal hatch).}
    \label{fig:Test3}
\end{figure*}

The purpose of this test was to quantify how far apart in velocity two identical features must be for \amoeba~to be able to reliably distinguish them. In this test, spectra with two identical Gaussian features and a signal-to-noise ratio of 5 were generated and \amoeba~was used to fit the spectra. The quality of the fits were judged to be `well' or `poorly fit' using the same criteria as the previous test. A fit was categorised as a `false negative' if one or no features were identified, and also if one of the two features' 99.95\% credibility intervals did not contain any of the correct values. A fit that identified three or more features was categorised as a `false positive'. This category was further subdivided into those where both, one or neither of the two features were `well fit'. The results of this test are illustrated in Fig. \ref{fig:Test3}. 

\begin{figure}
    \centering
    \includegraphics[trim={0cm 0cm 1cm 1cm}, clip=true, width=\linewidth]{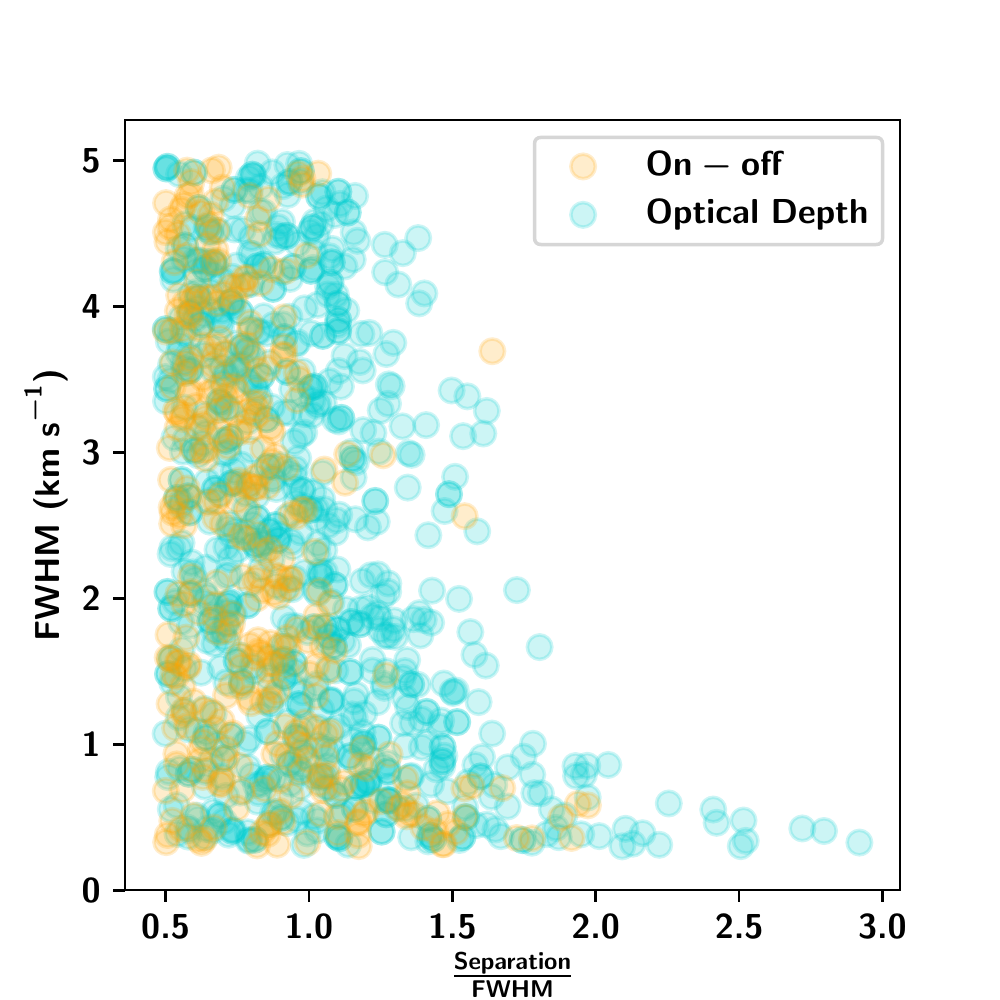}
    \caption{Separations and full widths at half-maximum for which \amoeba~returned a false negative result using on-off spectra (gold) and optical depth spectra (cyan), each with a velocity channel width of $0.1\,{\rm km\,s}^{-1}$.}
    \label{fig:FN}
\end{figure}

As expected, at separations less than the FWHM the majority of spectra returned a false negative result: the two features could not be distinguished. In both methods this false negative rate dropped off quickly, and $\approx 90\%$~of spectra were well fit at separations more than 2~times the FWHM for both the synthetic on-off spectra and optical depth spectra. However, this false negative rate is still high given that two identical features separated by the FWHM should be resolvable. In exploring the origin of this behaviour we identified that the features returning a false negative at these higher separations ($>1.5 \times \Delta v$) were skewed towards features with small FWHMs. In Fig. \ref{fig:FN} we plot the FWHM vs separation as a multiple of FWHM for all tests that returned a false negative result. This clearly shows a dramatic decrease in false negatives at separations greater than the FWHM, and that features with a greater separation only returned a false negative if they were quite narrow, i.e. $\Delta v < 1\,{\rm km\,s}^{-1}$, or equivalently $\Delta v < 10$~channels where the channel width is $0.1\,{\rm km\,s}^{-1}$. We attribute this to the lower number of data points across these features: the improvement to the likelihood distribution of fitting a narrow feature that only covers a few channels is much less than that of fitting a wide feature (see Eq. \ref{likelihood}). 

In a similar result to the previous test, the majority of fits categorised as `poorly fit' according to our criteria were not visually unreasonable, with typical examples appearing similar to those in Fig. \ref{fig:PF_2}. However, for both on-off and optical depth only spectra there were a significant number of false positive results. For synthetic on-off spectra, the majority of these did successfully identify the two correct features, but then detected additional features. For synthetic optical depth spectra approximately half of the total false positives successfully identified the two correct features (as well as one or more additional features), while the other half failed to identify either. In the vast majority of cases the spurious features had very low column density ($\log_{10} N_1\lesssim 10$) or optical depth ($|\tau_{\rm peak}| \lesssim 0.001$). We therefore suggest that the user take care when interpreting the identification of such low column density or optical depth features as they may indeed be false positives.

\section{Conclusions}\label{Sec:Conclusions}
    Here we have introduced \amoeba: an automated algorithm for the Bayesian Gaussian decomposition of hydroxyl spectra. \amoeba~takes a novel approach to the analysis of hydroxyl spectra by fitting on-off observations of all four ground state transitions simultaneously. This approach takes full advantage of the expected relationships between the features of these spectra, namely that components arising from the same ISM cloud should share the same centroid velocity and FWHM, and their peak values in each of the 8 spectra should be described uniquely by the column densities in the four levels of the ground-rotational state. It is also versatile in its ability to take as input both on-off spectra or optical depth spectra only.

We tested \amoeba~extensively, demonstrating the validity of its core Bayesian algorithm, its ability to recover the parameters of the synthetic spectra, as well as the impact of signal-to-noise ratio and feature separation on its ability to identify features. As summarised in Figs. \ref{fig:Test1} and \ref{fig:Test1_tau}, for 90\% of trial runs using unblended features with a signal-to-noise ratio of 5, \amoeba~is accurate to within 10\% when returning the centroid velocity, FWHM, excitation temperature, log column density and optical depth of synthetic spectra. When fitting `on-off' spectra it returns accurate parameters (i.e. where the `true' parameters are contained within the 99.95\% credibility interval, see Section \ref{sec:Test2} for a detailed explanation of this criterion) in over 90\% of trials above a signal-to-noise ratio of 2, and for optical depth spectra above a signal-to-noise level of 3 (with $\approx 80$\% accuracy at a signal-to-noise ratio of 2). In the case of two blended, identical features present in synthetic on-off spectra, \amoeba~is able to correctly identify both features in $>90$\% of trials at a separation of 1.5 times the FWHM, and at 1.75 times the FWHM for optical depth spectra. However, in the case of blended spectra with separations between 1 and 2 times the FWHM \amoeba~yielded false positives in a significant number of trials: $\approx$15\% for on-off spectra and $\approx$20\% for optical depth spectra.

\amoeba~is a powerful algorithm for the statistically robust Gaussian decomposition of hydroxyl spectra. By fitting all four ground-rotational state transitions simultaneously, \amoeba~is able to take full advantage of available data and produce self-consistent results with good precision. In fitting these spectra \amoeba~returns centroid velocity, FWHM, log column density in the lowest hyperfine level, and inverse excitation temperature in the 1612, 1665 and 1667 MHz transitions. These parameters, in conjunction with non-LTE molecular excitation code can then be used to find local parameters of the cloud, such as gas temperature, number density and local radiation field, etc. \amoeba~is completely automated and therefore an excellent tool for large-scale surveys in hydroxyl, such as THOR \citep{Beuther2016}.
\section*{Data Availability}
Data sharing is not applicable to this article as no new data were created or analyzed in this study.
\section*{Acknowledgements}

A.P. is the recipient of an Australian Government Research Training Program (RTP) stipend and tuition fee offset scholarship. J.R.D. is the recipient of an Australian Research Council (ARC) DECRA Fellowship (project number DE170101086), which partially supported this work. This work was performed on the OzSTAR national facility at Swinburne University of Technology. The OzSTAR program receives funding in part from the Astronomy National Collaborative Research Infrastructure Strategy (NCRIS) allocation provided by the Australian Government.

\clearpage
\appendix
    \section{Fitting a sample spectrum}\label{App:fitting}

Here we outline the process by which \amoeba~trials progressively more complex models using a set of synthetic on-off spectra (the same as in Fig. \ref{fig:test_vel}) containing 3 Gaussian-shaped profile features. The values of the parameters for these three Gaussian features are given in Table \ref{tab:App_spectra}. 

\begin{table}
    \centering
    \begin{tabular}{lccc}
    \hline
    \hline
    Parameter&Feature 1&Feature 2&Feature 3\\
    \hline
    $v$~(km\,s$^{-1}$)&-3&-1&3\\
    $\log_{10}\Delta v$~(km\,s$^{-1}$)&0.114&0.255&-0.097\\
    $\log_{10}N_1$~(cm$^{-2}$)&14&14.5&13.5\\
    $T_{\rm ex}^{-1} (1612)$~(K$^{-1}$)&0.2&-0.2&0.067\\
    $T_{\rm ex}^{-1} (1665)$~(K$^{-1}$)&0.067&0.1&0.2\\
    $T_{\rm ex}^{-1} (1667)$~(K$^{-1}$)&0.067&0.067&0.1\\
    \hline
    \hline
    \end{tabular}
    \caption{Parameters of the three Gaussian features in the synthetic spectra discussed in Section \ref{App:fitting}.}
    \label{tab:App_spectra}
\end{table}

This set of synthetic spectra was fit with \amoeba~a total of 26 times as a test of reliability. Though there was a systematic shift in the values found for some parameters, the standard deviation of the values found across these trials for each parameter was always significantly less ($90-95\%$) than the 68\%~confidence interval, and all were assessed as `well-fit' according to the definition used in Section \ref{Sec:Discussion}. During these 26 trials \amoeba~was set to only trial a single `extra' Gaussian feature after identifying one that did not improve the evidence. Therefore, included in the figures below will be two `over-fit' models for each velocity range. 

\begin{figure*}
    \centering
    \begin{tabular}{l}
    Step 1:\\
    \includegraphics[trim={0.6cm -.2cm 0.75cm 0.5cm}, clip=true, width=0.65\linewidth]{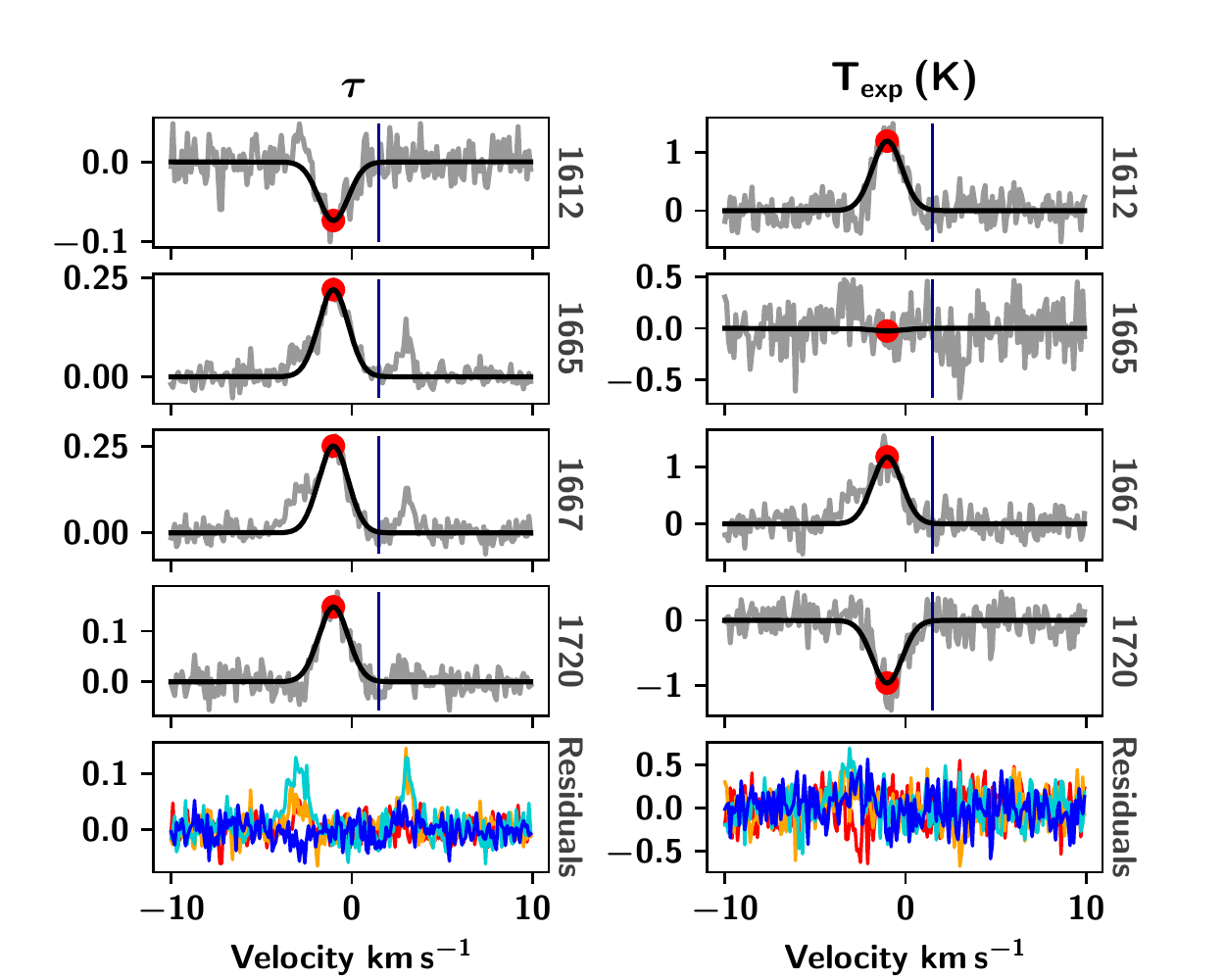}\\
    Step 2:\\
    \includegraphics[trim={0.6cm -.2cm 0.75cm 0.5cm}, clip=true, width=0.65\linewidth]{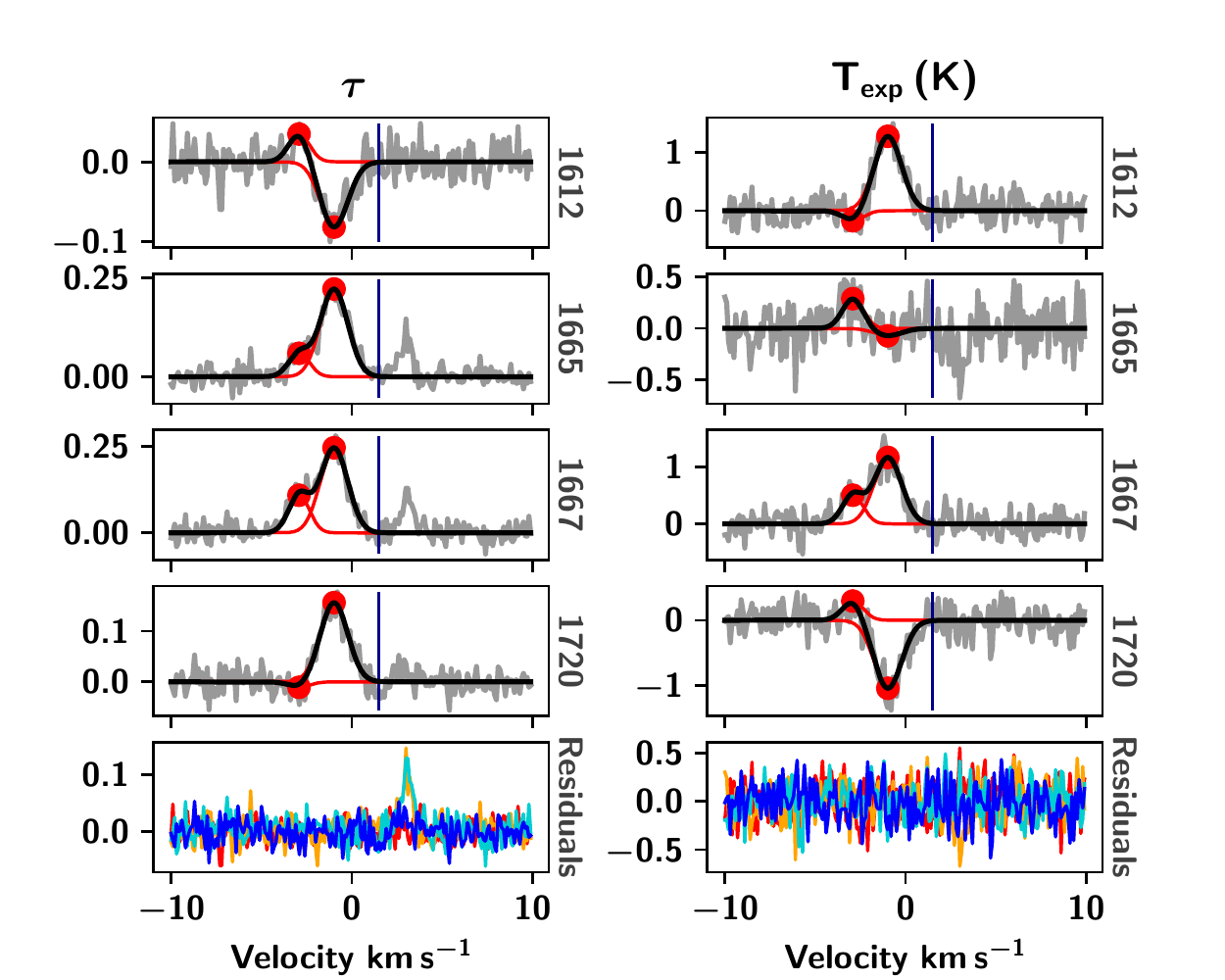}\\
    \end{tabular}
    \caption{}
    \label{fig:TEST_v1}
\end{figure*}

\begin{figure*}
    \centering
    \begin{tabular}{l}
    Step 3:\\
    \includegraphics[trim={0.6cm -.2cm 0.75cm 0.5cm}, clip=true, width=0.65\linewidth]{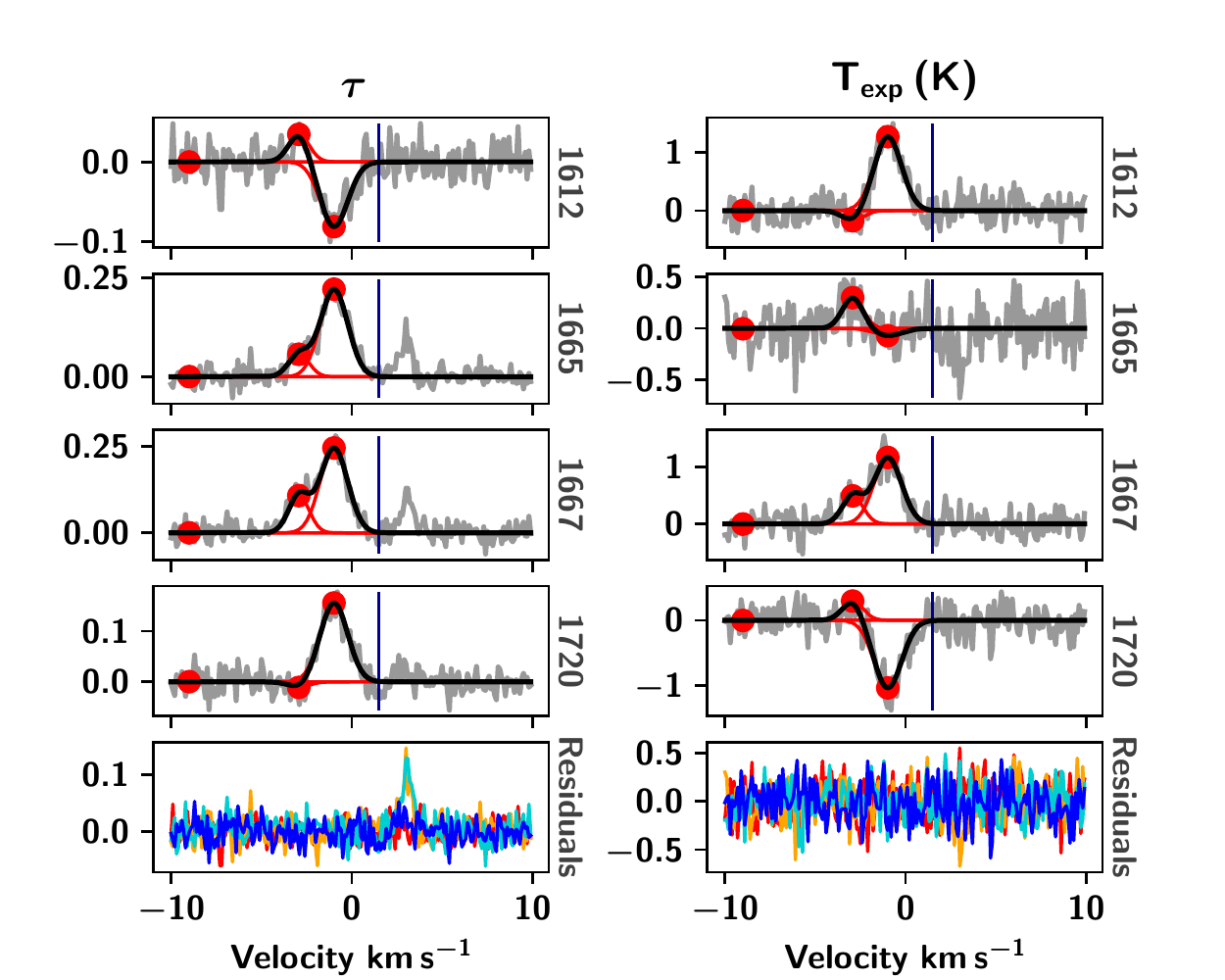}\\
    Step 4:\\
    \includegraphics[trim={0.6cm -.2cm 0.75cm 0.5cm}, clip=true, width=0.65\linewidth]{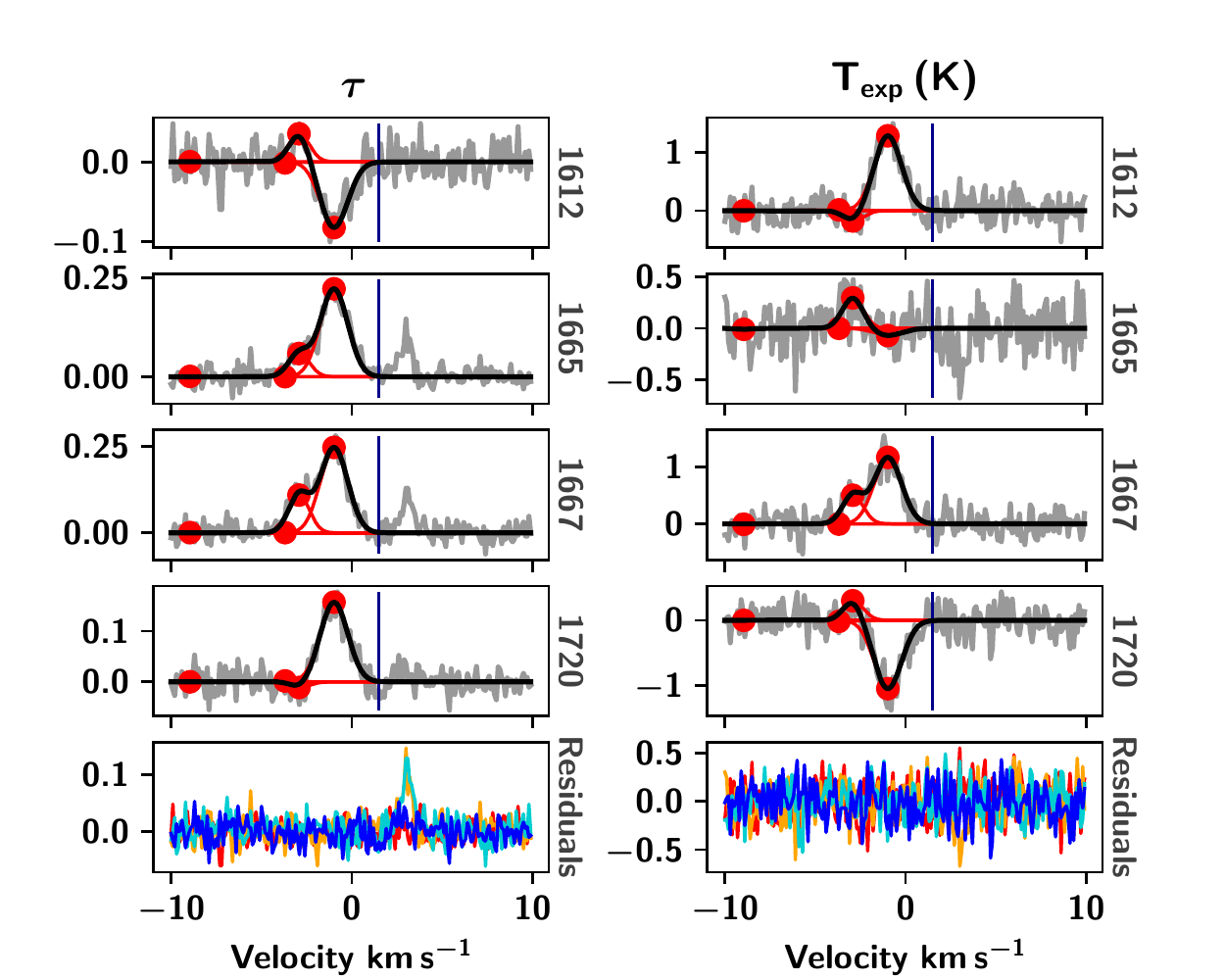}\\
    \end{tabular}
    \caption{}
    \label{fig:TEST_v2}
\end{figure*}

Each panel in Figures \ref{fig:TEST_v1} and \ref{fig:TEST_v2} shows the synthetic optical depth spectra at left and the expected brightness temperature at right in grey, and the velocity at which the spectra are cut with a vertical blue line. The `current' model is shown in black, with the individual Gaussian features that make up that model shown in red with their peaks indicated by a red circle. The residuals of the current model are shown in the bottom subplot, with 1612\,MHz in red, 1665\,MHz in gold, 1667\,MHz in cyan and 1720\,MHz in blue. The parameters of the models shown are outlined in Table \ref{tab:TEST_fit_params}.

\begin{table*}
    \centering
    \begin{tabular}{cccccccccc}
        \hline
        \hline
        Step&$v_{\rm min}$&$v_{\rm max}$&num. feat.&$v$&$\log_{10}\,\Delta v$&$\log_{10}\,N1$&$T_{\rm ex}^{-1}\,(1612)$&$T_{\rm ex}^{-1}\,(1665)$&$T_{\rm ex}^{-1}\,(1667)$\\
        \cline{2-3}\cline{5-6}\cline{8-10}
        &\multicolumn{2}{c}{km\,s$^{-1}$}&&\multicolumn{2}{c}{km\,s$^{-1}$}&cm$^{-2}$&\multicolumn{3}{c}{K$^{-1}$}\\
        \hline
        1&-10.0&-1.5&1&-1.02&0.29&14.5&-0.179&0.101&0.065\\
        \hline
        2&-10.0&-1.5&2&-2.92&0.13&14.0&0.201&0.067&0.067\\
        2&-10.0&-1.5&2&-0.98&0.25&14.5&-0.203&0.104&0.065\\
        \hline
        3&-10.0&-1.5&3&-9.00&-0.01&10.7&-1.695&0.734&0.314\\
        3&-10.0&-1.5&3&-2.93&0.13&14.0&0.202&0.065&0.067\\
        3&-10.0&-1.5&3&-0.99&0.26&14.5&-0.202&0.104&0.065\\
        \hline
        4&-10.0&-1.5&4&-8.95&0.01&11.1&0.604&0.675&0.328\\
        4&-10.0&-1.5&4&-3.69&0.04&11.5&-2.410&0.087&0.155\\
        4&-10.0&-1.5&4&-2.92&0.13&14.0&0.203&0.066&0.067\\
        4&-10.0&-1.5&4&-0.98&0.25&14.5&-0.204&0.104&0.065\\
        \hline
        5&-1.5&10.0&1&3.01&-0.10&13.5&0.016&0.180&0.099\\
        \hline
        5&-1.5&10.0&2&3.00&-0.07&13.5&0.029&0.181&0.097\\
        5&-1.5&10.0&2&3.75&-0.05&11.6&-2.062&0.219&0.291\\
        \hline
        7&-1.5&10.0&3&3.02&-0.07&13.5&0.026&0.184&0.097\\
        7&-1.5&10.0&3&3.47&-0.03&11.1&-5.556&-0.203&0.412\\
        7&-1.5&10.0&3&9.02&-0.02&10.9&-13.121&0.334&0.226\\
        \hline
        Final&-10.0&10.0&3&-2.92&0.13&14.0&0.201&0.067&0.067\\
        Final&-10.0&10.0&3&-0.98&0.25&14.5&-0.203&0.104&0.065\\
        Final&-10.0&10.0&3&3.01&-0.10&13.5&0.016&0.180&0.099\\
        \hline
        \hline
    \end{tabular}
    \caption{Parameters of the Gaussian features identified by \amoeba~in each step of the Gaussian decomposition process illustrated in Figures \ref{fig:TEST_v1} and \ref{fig:TEST_v2}.}
    \label{tab:TEST_fit_params}
\end{table*}

\begin{figure*}
    \centering
    \begin{tabular}{l}
    Step 5:\\
    \includegraphics[trim={0.6cm -.2cm 0.75cm 0.5cm}, clip=true, width=0.65\linewidth]{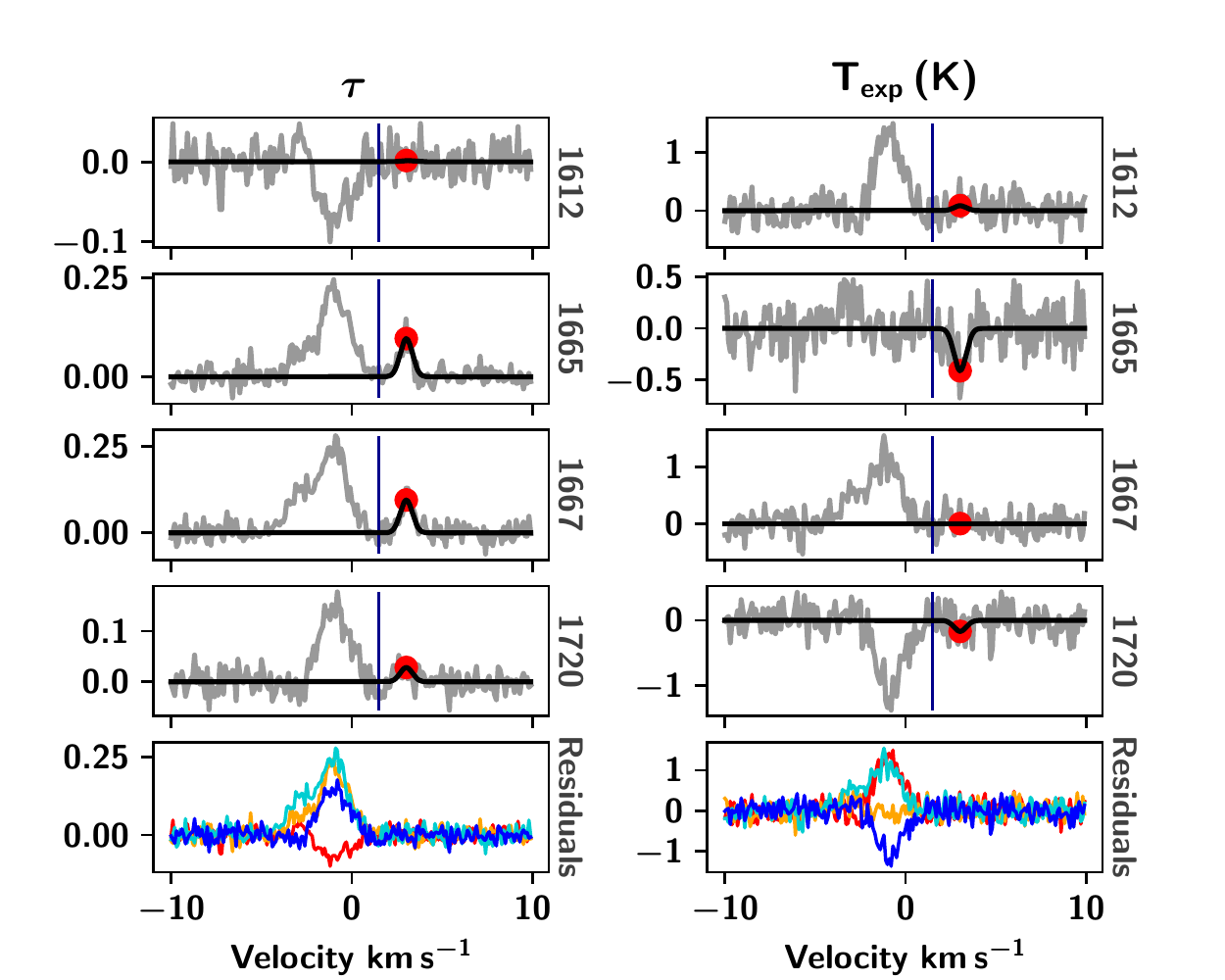}\\
    Step 6:\\
    \includegraphics[trim={0.6cm -.2cm 0.75cm 0.5cm}, clip=true, width=0.65\linewidth]{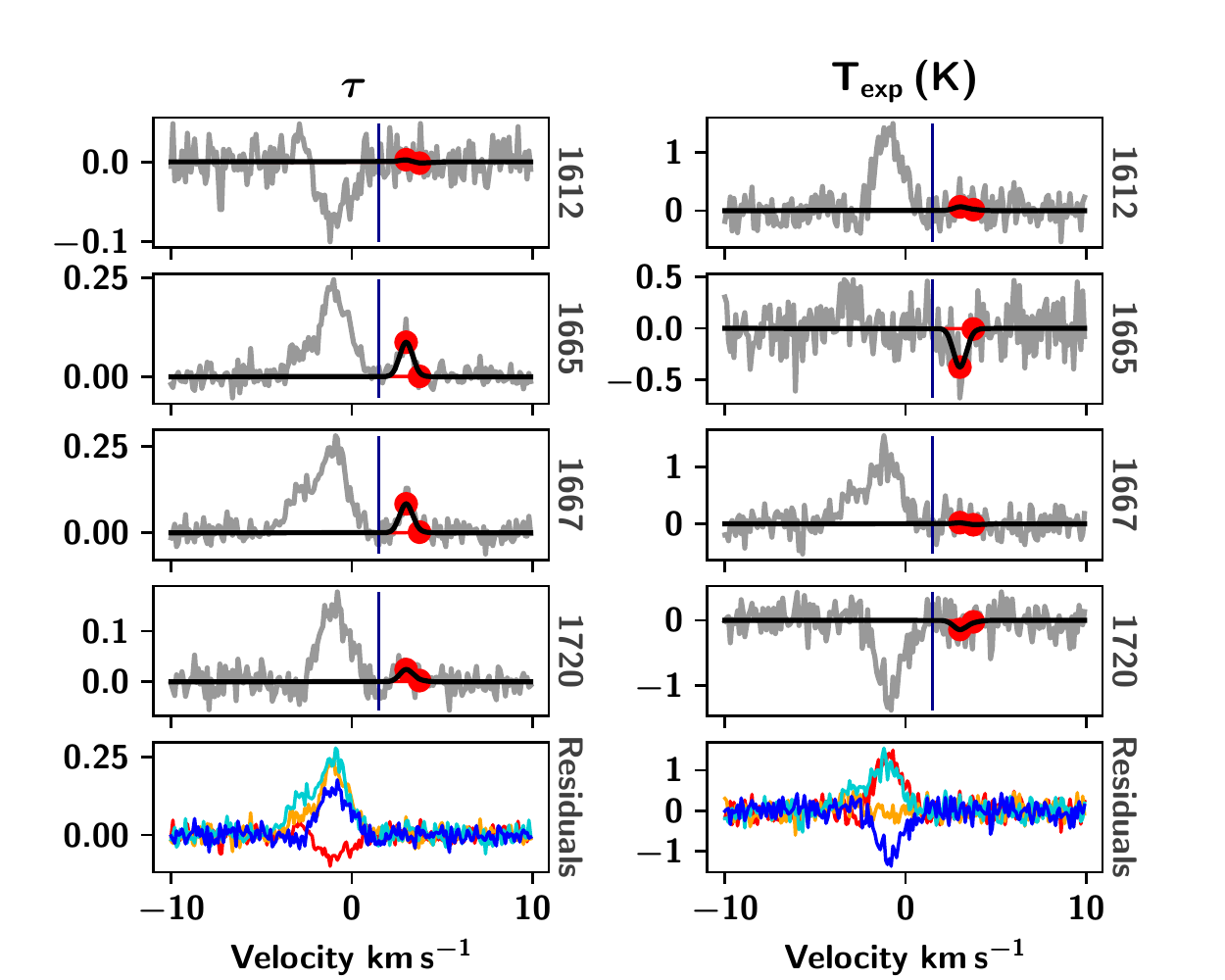}\\
    \end{tabular}
    \caption{}
    \label{fig:TEST_v3}
\end{figure*}

\begin{figure*}
    \centering
    \begin{tabular}{l}
    Step 7:\\
    \includegraphics[trim={0.6cm -.2cm 0.75cm 0.5cm}, clip=true, width=0.65\linewidth]{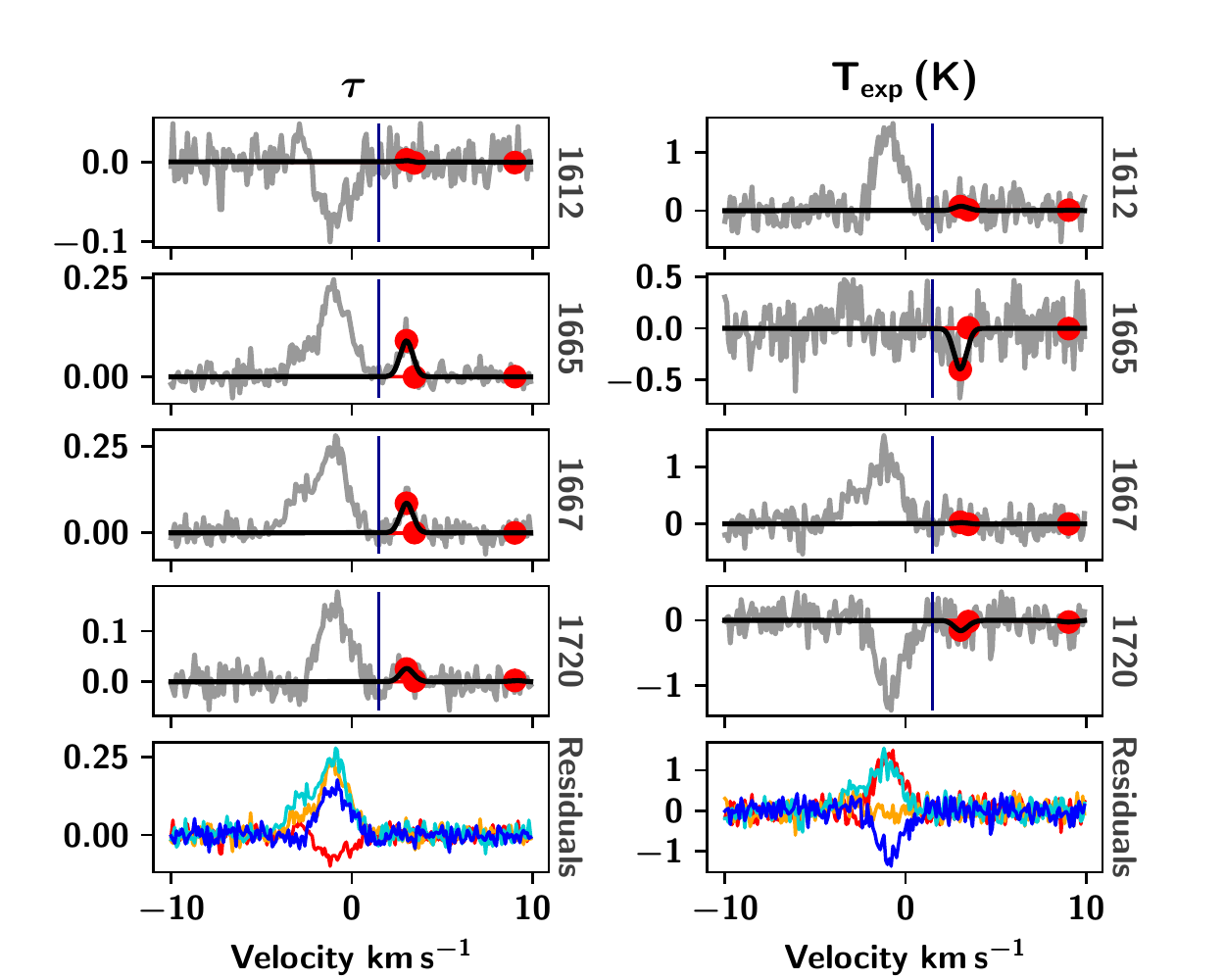}\\
    Final:\\
    \includegraphics[trim={0.6cm -.2cm 0.75cm 0.5cm}, clip=true, width=0.65\linewidth]{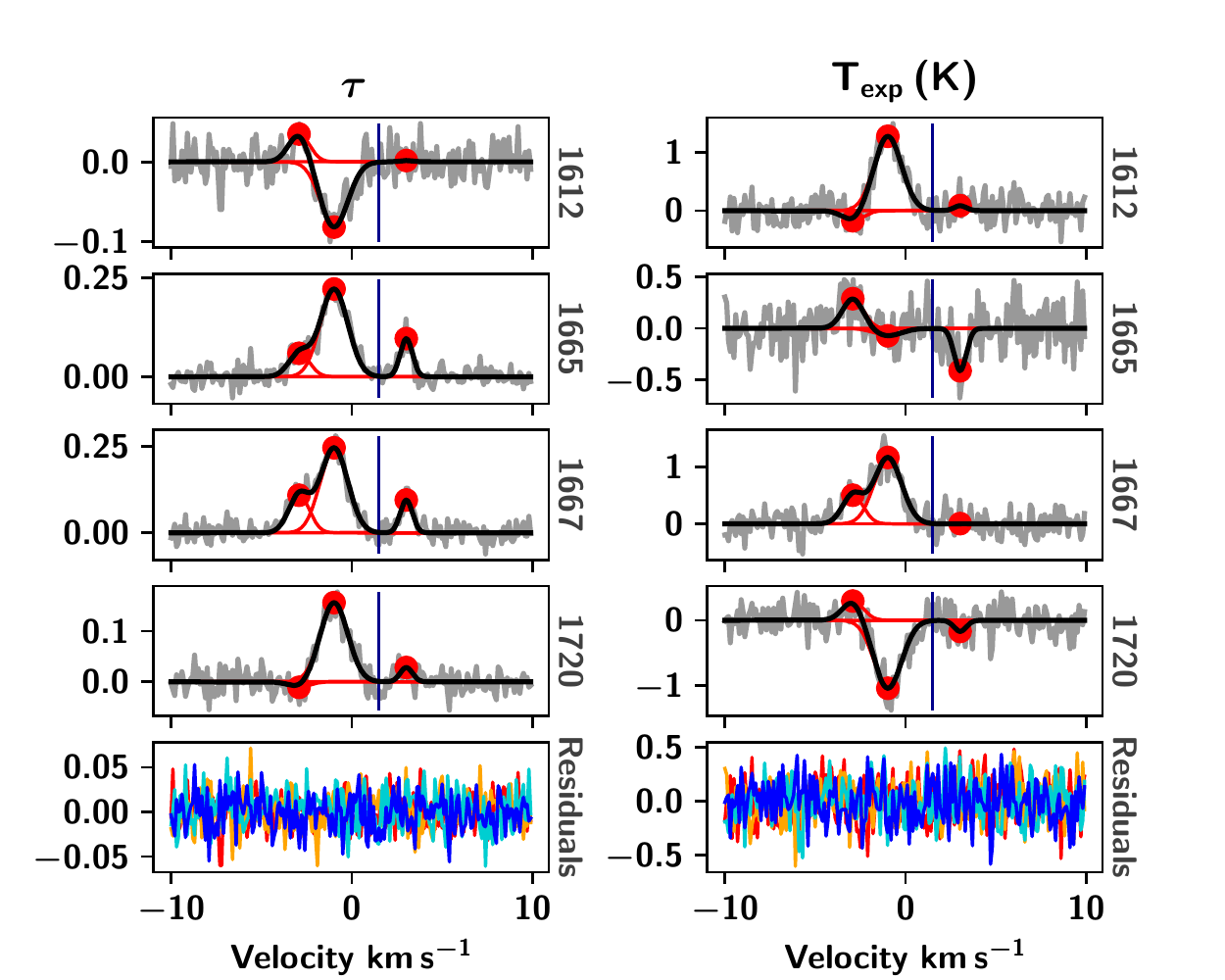}\\
    \end{tabular}
    \caption{}
    \label{fig:TEST_v4}
\end{figure*}

After fitting a null model to the spectra from -10 to 1.5 km\,s$^{-1}$~(resulting in $\log_{\rm e}{\rm null~evidence}=-725$). \amoeba~then sampled the posterior probability distribution of a model with a single Gaussian component (`Step 1' in Fig. \ref{fig:TEST_v1}). This model had $\log_{\rm e}{\rm evidence}=1019$~and was therefore accepted over the null model. \amoeba~then sampled the probability distribution of a model with two Gaussian components (`Step 2' in Fig. \ref{fig:TEST_v1}). This model had $\log_{\rm e}{\rm evidence}=1087$~and was therefore accepted over the model with a single Gaussian. This model was found to have the highest evidence of those trialled and was accepted as the final model for this velocity range, and is included in the `Final' plot in Fig. \ref{fig:TEST_v4}.

After accepting the model with two Gaussian components, \amoeba~then sampled the posterior probability distribution of a model with three Gaussian components (`Step 3' in Fig. \ref{fig:TEST_v2}). In this case \amoeba~identified the same two components it had accepted in the previous model, as well as a third component. For the sake of illustration, an arbitrary example of this fit from the 26 trials is shown in Fig. \ref{fig:TEST_v2}, but the median position in parameter space of the converged Markov chains varied widely (the third Gaussian was placed anywhere from -9 to -1\,km\,s$^{-1}$). However, the average measured evidence of this model ($\log_{\rm e}{\rm evidence}=1050$) did not vary significantly, and the model was rejected each time. The posterior probability distribution of a model with 4 Gaussian components was then sampled (`Step 4' in Fig. \ref{fig:TEST_v2}). Again, the components from the previously accepted model were identified, along with two additional features. While the median positions of these extra features again varied widely, the evidence was relatively constant ($\log_{\rm e}{\rm evidence}=1000$) and the model was rejected each time.

\amoeba~then moved to the next velocity range from 1.5 to 10 km\,s$^{-1}$~which had $\log_{\rm e}{\rm null~evidence}=807$. \amoeba~then sampled the posterior probability distribution of progressively more complex models in the same way as for the previous velocity range, and the models represented by the median positions of the converged Markov chains are illustrated in `Step 5' and `Step 6' in Fig. \ref{fig:TEST_v3} and `Step 7' in Fig. \ref{fig:TEST_v4}. The model with a single Gaussian component was found to have $\log_{\rm e}{\rm evidence}=834$~and was therefore accepted over the null model. When the posterior probability distributions of models with two and three Gaussian components were sampled, the median position of the converged Markov chains varied, but the $\log_{\rm e}{\rm evidence}$~values were relatively constant at 806 and 764, respectively, and these models were rejected in all 26 trials. The lower panel of Fig. \ref{fig:TEST_v4} (`Final') then shows the final model accepted by \amoeba.

\section{Typical corner plots}

\begin{figure*}
    \centering
    \includegraphics[trim={0cm 2cm 13cm 2cm}, clip=true, width=\linewidth]{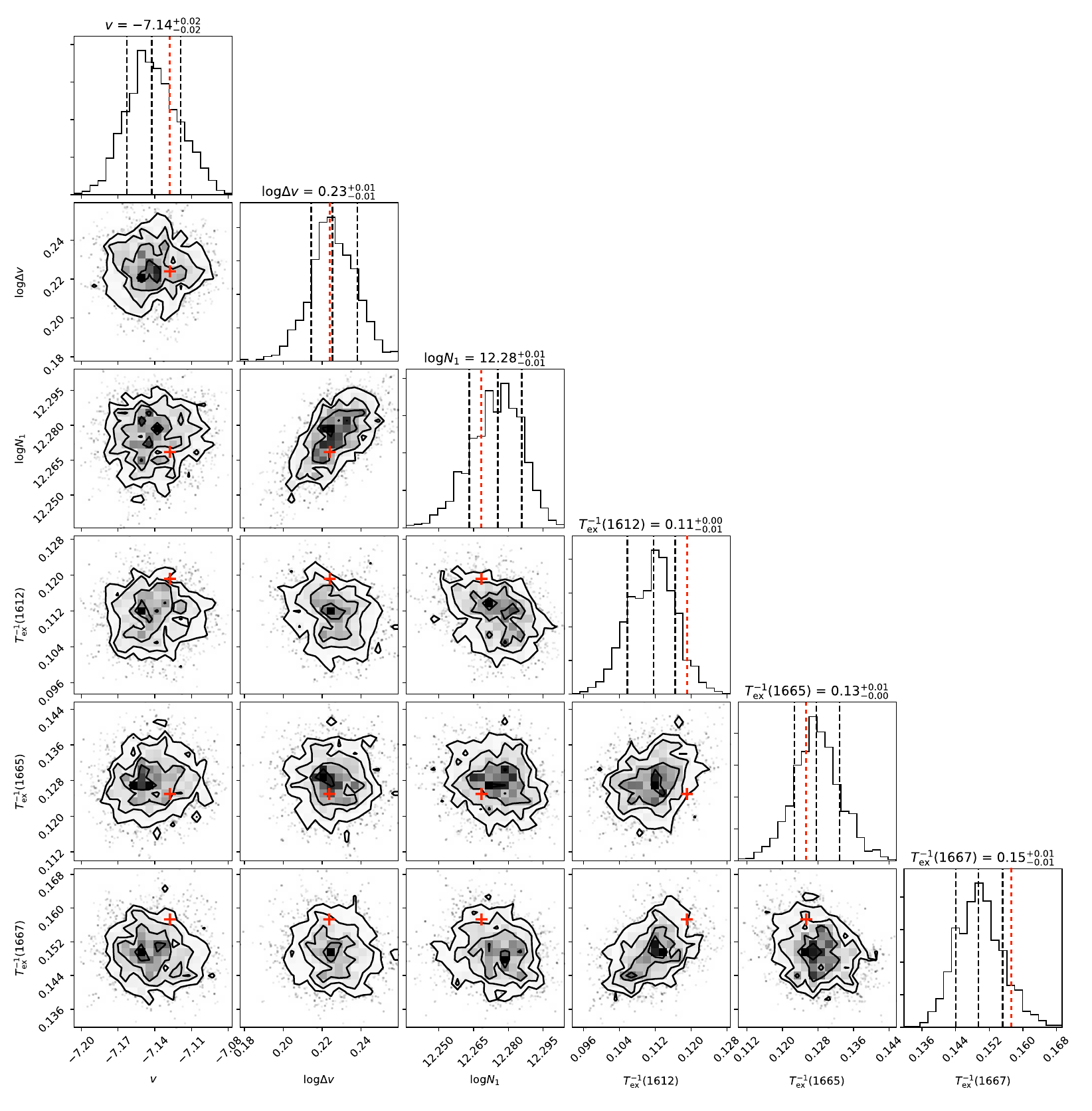}
    \caption{A typical corner plot from a single test of \amoeba~on a set of synthetic on-off spectra with a single Gaussian feature and a signal-to-noise ratio of 5. The true parameter values are indicated with red dotted lines on each histogram and with red crosses on each scatter plot. Only minor correlations between some pairs of parameters are seen in this example. This plot was generated using the Corner package in \textsc{python} \citep{Foreman-Mackey2016}.}
    \label{fig:corner_Texp}
\end{figure*}

\begin{figure*}
    \centering
    \includegraphics[trim={0cm 2cm 13cm 2cm}, clip=true, width=\linewidth]{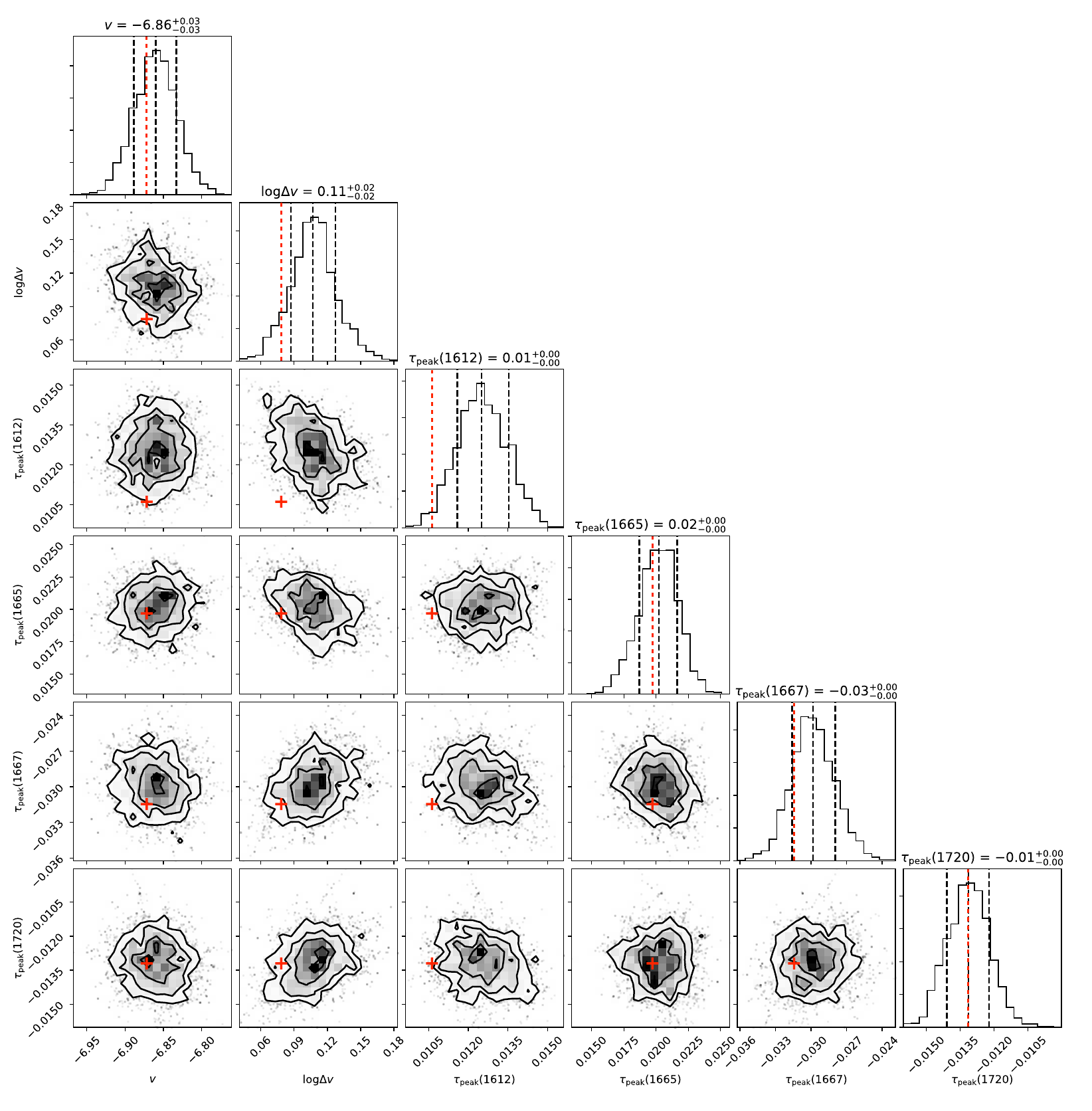}
    \caption{Same as Fig. \ref{fig:corner_Texp} but for synthetic optical depth only spectra. No significant correlations are seen in this example.}
    \label{fig:corner_tau}
\end{figure*}

\bibliography{bib}{}
\bibliographystyle{aasjournal}



\end{document}